\documentclass[iop, twocolumn, numberedappendix]{emulateapj-rtx4}
\usepackage{comment}
\usepackage{longtable}
\usepackage{placeins} % I have an inordinate amount of floats that causes Latex to break.  making a \FloatBarrier is the only way I can think to solve this issue
\usepackage{amsmath}
\usepackage{color}
\usepackage{pdflscape}
\usepackage{pdfpages}
\usepackage{epsfig,graphicx}
\usepackage{epstopdf}
\usepackage[urlcolor=blue]{hyperref}

\def\etal{{et~al.}}
\def\kms{{\hbox{km s$^{-1}$}}}

\shorttitle{Intrinsic HI and OH Absorption in Compact Radio Sources} 
\shortauthors{Grasha \etal}

\begin{document}
\title{A Search for Intrinsic HI 21~cm and OH 18~cm Absorption Toward Compact Radio Sources}    

\author{Kathryn Grasha\altaffilmark{1,2,3}, Jeremy Darling\altaffilmark{3}, Alberto Bolatto\altaffilmark{4}, Adam K. Leroy\altaffilmark{5}, John T. Stocke\altaffilmark{3}}
\altaffiltext{1}{Research School of Astronomy and Astrophysics, Australian National University, Canberra, ACT 2611, Australia; kathryn.grasha@anu.edu.au}
\altaffiltext{2}{ARC Centre of Excellence for All Sky Astrophysics in 3 Dimensions (ASTRO 3D), Australia}
\altaffiltext{3}{Center for Astrophysics and Space Astronomy, Department of Astrophysical and Planetary Sciences, University of Colorado, 389 UCB, Boulder, CO 80309-0389, USA}
\altaffiltext{4}{Department of Astronomy, University of Maryland, College Park, MD 20742-2421, USA}
\altaffiltext{5}{Department of Astronomy, The Ohio State University, McPherson Laboratory, 140 West 18th Avenue, Columbus, OH 43210-1173, USA}

\begin{abstract}
We present the results of a large search for intrinsic HI 21~cm and OH
18~cm absorption in 145 compact radio sources in the redshift range $0.02<
z <3.8$ with the Green Bank Telescope. We re-detect HI 21~cm 
absorption toward six known absorption systems but detect no new HI or OH
absorption in 102 interference-free sources.  
79 sources have not previously been observed for HI 21~cm
absorption. We recover a mean optical depth limit of
$\tau_{3\sigma}<0.023$ for all the non-detections in the survey.  
Our results do not support
the high intrinsic absorption rates found by previous studies in compact
radio sources at low redshift. Our results do, however, support the hypothesis
proposed by Curran \etal\ (2008) that high ultraviolet (UV) 
luminosity active galactic nuclei (AGN) do not show intrinsic HI 21~cm
absorption, confirming a threshold of $L_{\rm UV} =
10^{23}$~W~Hz$^{-1}$, above which our intrinsic absorption fraction is zero
(54 sources). The exact nature of the UV luminosity effect on HI absorption systems remains ambiguous. We additionally find no statistical correlation between the 1.4~GHz radio luminosity or the source size and the 21~cm absorption detection rate. 
We attribute the lack of intrinsic absorption in our survey to the UV luminosity effect
caused by an optical selection bias and a decreased column density
sensitivity with increasing redshift due to lower radio continuum flux
densities, high radio frequency interference, and higher telescope system temperatures at low frequencies.  
\end{abstract}

\section{Introduction}\label{sec:intro}
Many compact radio sources are likely to be young and/or confined radio jets. It is likely that these very young and emerging jets are still contained within their host galaxy, potentially providing favorable sites to observe the interactions between the young, compact jets and the intrinsic gaseous interstellar medium (ISM) near the central active galactic nuclei (AGN) within their galaxies \citep{odea97, owsianik98, readhead96}. 

One of the best observational tracers of the intrinsic ISM is with observations of the cold neutral medium (CNM) as traced by the hyperfine spin-flip transition of neutral hydrogen, the HI 21~cm line. When cold neutral hydrogen (T $\sim$ 100~K) is present, absorption of the 21~cm line against these compact radio sources gives the ability to detect neutral hydrogen absorption systems at any redshift. The 21~cm absorption line \textbf{could} thus be an invaluable probe of the kinematics and morphology of the neutral gas in the nuclear regions of galaxies.  Studies of HI quite often employ the use of compact radio sources as the best targets to detect 21~cm absorbers as the jets are still contained within their host galaxies, providing insight to the possible feedback interactions of these young jets with the neutral ambient medium.

Since the seminal work of \citet{roberts70} that identified extragalactic associated HI 21~cm absorption in NGC~5128, a large number of HI 21~cm absorption studies have been carried out in an attempt to provide a better understanding of the properties of neutral gas, especially within compact radio source host galaxies. The discovery of blueshifted HI absorption systems at large velocities \citep[$\gtrsim - 1000$~\kms;][]{ver03} with respect to the systemic optical redshift of the galaxy hosting the compact radio source is indicative of jet-driven outflows of neutral gas. This signifies a unique avenue in which to study the fueling of the AGN as well as the effects of AGN feedback and its interaction with the environment and gas kinematics of the host galaxy.

The majority of HI 21~cm absorption systems have high column density and are damped Ly$\alpha$ (DLA; $N_{\rm HI} \geq 2\times 10^{20}$~cm$^{-2}$) systems. 
The vast majority of 21~cm absorption systems have low redshifts, $z\lesssim1$ \citep[for a compiled literature list, see][]{curran16} but some of intrinsic absorbers do lie above $z\gtrsim1$ \citep{uson91, moore99, ishwara03, curran13, aditya17, aditya18a}. 
Previous surveys have identified a few commonalities exhibited in intrinsic HI 21~cm absorbers:  

(1) High detection rates, of order 30-50\%, are found in surveys of 
compact radio sources at redshift $z\lesssim1$ \citep[e.g.,][]{ver03, pih03, gupta06, gereb15, maccagni17} whereas the detection rates in higher
redshift surveys are significantly lower \citep[e.g.,][]{kanekarchengular03, curran08, curran11, aditya16, aditya17, curran17, moss17}. 
The deficiency of HI 21~cm absorbers at high redshift could arise from detection biases in sample selection or could signify physical changes in the reservoirs of cold neutral gas, AGN activity, or radio emission at earlier cosmic time  \citep[although neutral gas fractions are generally expected to increase with redshift;][and references therein]{rhee18}. 

(2) No intrinsic 21~cm absorption system has ever been detected toward objects with an ultraviolet (UV) luminosity 
$L_{\rm UV} > 10^{23}$~W~Hz$^{-1}$ \citep{curran08, curran11, curran17} at any redshift. This luminosity is indicative of a UV bright, unobscured quasar with minimal extinction along the line of sight. 
While the exact nature of the UV luminosity's effect on neutral gas 
remains somewhat ambiguous, the observed threshold likely represents an analogue to the AGN 
``proximity effect'' seen in UV absorption line studies \citep{baj88}. The effect on the HI 21 cm absorption line 
could be the result of ionization of gas from the UV radiation \citep{curran19}, an increase in the spin temperature of the cold neutral medium \citep[e.g., ][]{kanekarchengular03}, effectively decreasing the observed line optical depth for a given column density, or simply identifying systems that are unextincted and have little gas and dust along the line of sight.  

(3) High redshift, optically bright AGN are most likely to have spectroscopic redshifts. 
Most radio sources with spectroscopic redshifts beyond $z\sim1.2$ have UV luminosities in excess of $L_{\rm UV}>10^{23}$~W~Hz$^{-1}$. 
The UV luminosity threshold effect is therefore especially
problematic for high redshift 21~cm absorption line detection because optical spectroscopic redshifts of AGN rely on 
rest-frame UV emission lines, biasing target selection toward the brightest UV sources that have never been detected in 21~cm absorption. 

Another gas tracer that may be associated with neutral hydrogen absorption is the hydroxyl molecule (OH). OH shows absorption, emission, and maser activity in four hyperfine and lambda-doubled transitions at 18~cm. Cospatial HI and OH lines allow one to constrain the cosmic evolution of several physical constants: the electron-proton mass ratio $\mu \equiv m_p/m_e$, the fine structure constant $\alpha = e^2/\hbar c$, and the proton g-factor $g_p$
\citep{chengalur03, darling03, darling04, kanekar04, kanekar05, tzanavaris05}. The fine structure constant $\alpha$ parametrizes the strength of electromagnetic interactions, and some quasar studies suggest a decrease in $\alpha$ in the early universe \citep[e.g., ][]{cowie95, murphy01, darling04, tzanavaris07, kanekar06, srianand10, rahmani12}. Changes in the physical constants over cosmic time will cause spectral lines to shift from their relative measured values at present-day and have implications for physics and cosmology. 

Redshifted OH systems are rarer than HI 21~cm absorption systems and are always found in systems that are detected in 21~cm absorption as well. Five OH 18~cm absorption line systems are known to date at
$z>0.2$ \citep{chengalur99, kanekar03, darling04, kanekar05, kanekar12}, and none lie beyond $z=1$. 
Compact radio sources still confined within the host galaxy are the perfect environment to search for molecular systems, producing optimal conditions to observed OH absorption lines, especially the conjugate\footnote{The satellite OH conjugate lines (1612 and 1720~MHz) are a rare astrophysical phenomenon where the two lines have the same shape but opposite signs, arising due a population anti-inversion mechanism resulting from quantum mechanical selection rules \citep{elitzur92, vanlang95, darling04, kanekar18}.} OH lines, systems in which OH gas complexes become sources of coupled weak maser emission and non-thermal absorption. Large surveys over a broad redshift range are required to detect the rare coincidence of HI 21~cm and OH 18~cm absorption lines.

In this paper, we perform a large survey for intrinsic HI 21~cm and OH 18~cm
absorption in the host galaxies of 145 compact radio sources, including
GHz-Peaked Spectrum (GPS), Compact Steep-Spectrum (CSS), Compact Symmetric Objects (CSOs), and Flat-Spectrum Radio Sources (FSRSs; see Section~\ref{sec:sample} for further discussion of these classifications), using the Green Bank Telescope\footnote{The
  National Radio Astronomy Observatory is a facility of the National
  Science Foundation operated under cooperative agreement by Associated
  Universities, Inc.} (GBT) selected across the redshift range ($0.02< z <3.8$) in an attempt to increase the number of known absorbers, especially at high redshift. In addition to the results of our search
for intrinsic HI and OH absorption, we examine several biases that influence
the detection of intrinsic absorption, including redshift selection, UV
luminosity, radio luminosity, radio source size, covering factor, and spin
temperature. We explore the possibility that the nature of our
high redshift selection of radio sources biases the results against
detection of intrinsic 21~cm absorption. Previous 21~cm absorption studies of the same
biases we investigate here were performed on a heterogeneous compilation of
previous known intrinsic HI absorption with a wide range of source sizes that were typically located at low redshift \citep[e.g.,][]{ver03, gupta06, gereb14}. These results are arguably not unbiased because non-detections have generally not been uniformly published, which will skew the detection rates. We use this large survey to test the UV luminosity threshold hypothesis and the high rate of intrinsic HI 21 cm absorption toward compact radio sources. 

The paper is outlined as follows. The compact radio source sample selection is described in Section~\ref{sec:sample} and their observations are described in Section~\ref{sec:observations}. The
results and comparison to previous 21~cm absorption searches are presented in
Section~\ref{sec:results}. Section~\ref{sec:analysis} presents the
analysis of the HI detection rate with the
radio and UV luminosities. We discuss the findings in Section~\ref{sec:discussion} and we summarize the results and conclusions in Section~\ref{sec:conclusions}.   

Throughout this paper, we adopt a flat $\Lambda$CDM cosmology
with $\Omega_{\rm m} = 0.27$, $\Omega_{\Lambda} =0.73$, and $H_\circ =
71$~\kms~Mpc$^{-1}$ \citep{komatsu11}.  All quantities obtained from the
literature are recalculated (if necessary) for this cosmology.

\section{Sample Selection}\label{sec:sample}
Compact radio sources are often classified by their morphology and radio spectral index, providing insight into the properties and nature of these sources. GHz-Peaked Spectrum (GPS) sources are intrinsically small \citep[i.e., not foreshortened by projection effects;][]{fanti90} and show a low frequency turn-over in their spectra, mainly attributed to synchrotron self-absorption and free-free (thermal bremsstrahlung) absorption \citep{jones74, menon83, odea97}. Compact Steep-Spectrum (CSS) sources show steeper power-law spectra than is typically seen in radio galaxies \citep{peacock82}. Compact Symmetric Objects (CSOs) show a compact double-lobed structure with advance speeds, indicative of young jets \citep{wilkinson94,owsianik98}. Flat-Spectrum Radio Sources (FSRSs) are characterized with a double-peaked synchrotron/Compton spectral energy distribution, possibly associated with a blazar or with the compact core of a radio galaxy \citep{fugmann88, molina12}. These classifications are not disjoint in the literature; for example, most CSOs are also classified as GPS sources \citep{xiang05}.  

Candidate intrinsic HI 21 cm absorbers in this survey include nearly all known GPS sources, CSS sources, and CSOs (ca. 2005) with $\delta>-35^\circ$ \citep{spencer89, fanti90, devries97, morganti97, odea97, peck00, xiang05} for a total of 60 objects (34 GPS sources, 29 CSS sources, and 7 CSOs, with some overlap between the classifications). The FSRSs were selected from \citet{white92} with $S>0.3$~Jy at 780~MHz, adding an additional 85 sources to the sample selection.  

\begin{figure}
\epsscale{1.2}
\plotone{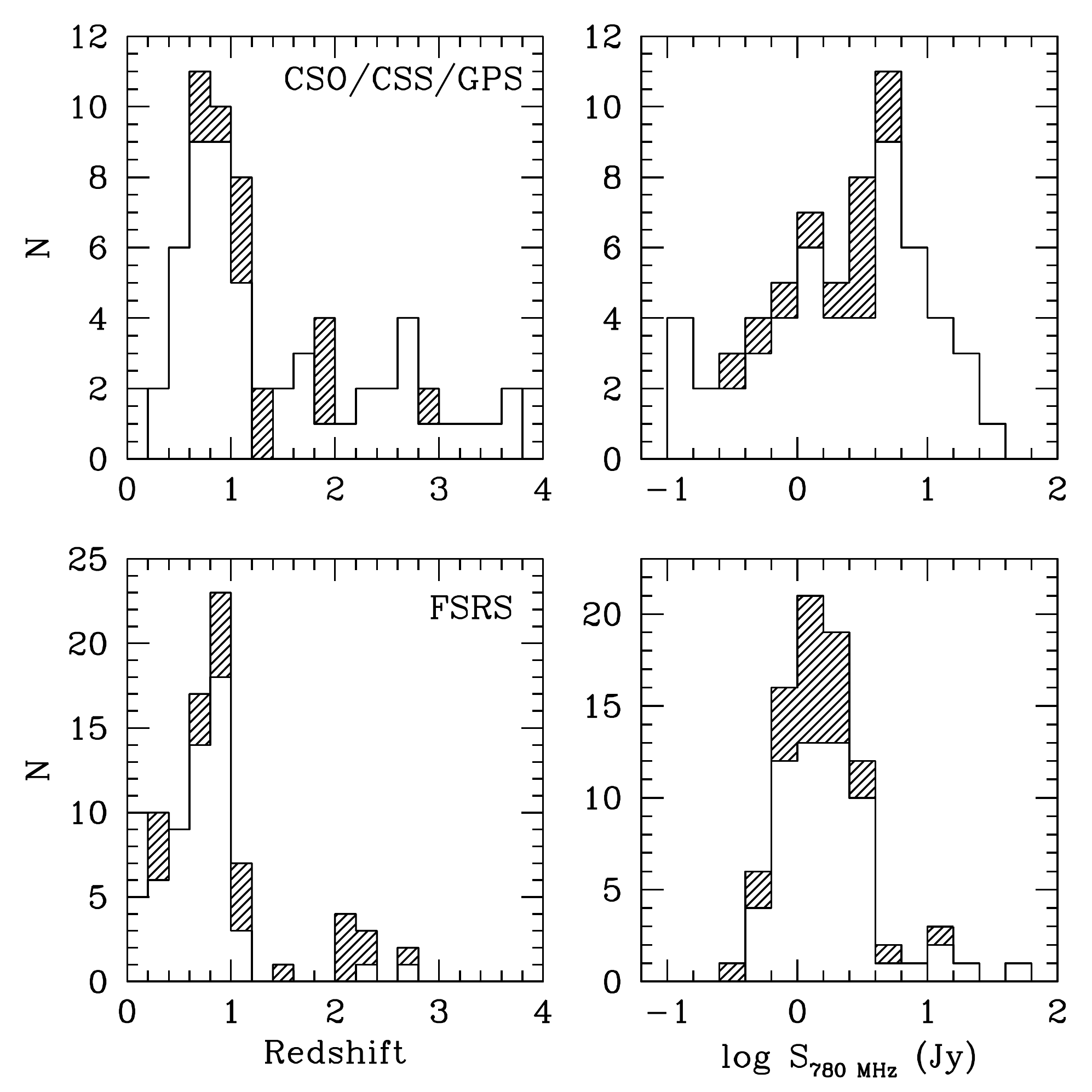}
\caption{Redshift (left) and continuum flux density at the expected redshifted 21 cm line frequency (right) for the CSO, CSS, and GPS sources (top row) and the FSRSs (bottom row) in our sample. The hatched cells show the objects that could not be observed for spectral lines due to RFI. 
\label{fig:sources}}
\end{figure}

In total, the sample includes 145 objects with known optical redshifts $0 < z < 4$, with the vast majority of the sample at redshifts less than $z\sim1$. Figure \ref{fig:sources} shows the redshift and the continuum flux 
  density at the expected redshifted line frequency of the two samples (compact and flat-spectrum sources; Section~\ref{sec:flux}).   
Table \ref{tab:table1}
lists the sources, their radio
classifications (GPS, CSS, CSO, and FSRS), total integration time for each reduced spectrum.

\section{Observations and Data Reduction}\label{sec:observations}
From September 2004 to August 2005, we observed 85 FSRSs for intrinsic absorption in the HI 21~cm 
line (1420.405752~MHz), 
and 81 in the four 18~cm OH lines at
1612.231, 1665.4018, 1667.35903, and 1720.5300~MHz simultaneously (GBT program 04C--018). 
We additionally observed 60 GPS, CSS, and CSO radio sources for intrinsic 
absorption from December 2005 to May 2006 (GBT program 06A--044): 
 59 sources were observed in the HI 21~cm line,
and 44 sources were observed in the four OH 18~cm lines. One source,
PKS~2135$-$209, was only observed in the OH lines.  
All observations were conducted 
using a 5-minute position-switched mode in two linear polarizations with
9-level sampling, 3- to 15-sec records, and a 50~MHz bandwidth
centered at the redshift of the radio source host galaxy.  
Each bandwidth contains 4096 spectral channels with a width of 12.2 kHz per channel (a channel width of 3.86~\kms\ at $z=0.5$).  
A calibration diode signal was injected for half of each record. Total
on-source integration times were typically 30 minutes for the GPS, CSS, and
CSO sources and 5 minutes for the FSRSs, listed in Table~\ref{tab:table1}. 

Each off-source-flattened and diode-calibrated spectral record was inspected
and flagged for radio frequency interference (RFI).  All records were then
averaged in time and polarization, Hanning smoothed, and 
baseline-flattened using a polynomial fit, typically of third
order with a final post-smoothing spectral resolution of 24.4~kHz per channel.  
All data reduction was performed using
GBTIDL\footnote{GBTIDL (\url{http://gbtidl.nrao.edu/}) is the data
  reduction package produced by NRAO and written in the IDL language for
  the reduction of GBT data.}.  
Final mean spectra were inspected for absorption lines within a few thousand km~s$^{-1}$ 
of the optical redshift of the host galaxy. The observable velocity span was typically determined by the RFI
conditions rather than the spectral bandwidth.

\section{Results}\label{sec:results} 
The nature of 36 sources (25\% of the total sample) remains ambiguous due to unrecoverable contamination of RFI, a pernicious impediment to the detection of new absorption systems. The remaining 108 sources have the majority of their spectral range free from RFI and are able to be searched for HI or OH absorption. 

Table \ref{tab:HI} lists the RFI-free velocity range searched
for HI 21~cm absorption toward each radio source, the measured rms noise (away
from RFI and spectral lines)
in the mean spectra, the continuum flux density at the expected
frequency of the redshifted spectral line obtained from power-law
fits to extant continuum measurements in the literature (see Section~\ref{sec:flux}), HI column densities (or upper limits toward radio sources
not detected in 21~cm absorption), and the projected size of
each radio source as obtained from literature searches (see Section \ref{sec:size}).  

Table \ref{tab:OH} lists the properties of the OH 18~cm line observations toward a subset of the
sample.  The 1665 and 1667~MHz lines are both observed, but we only list the
properties of the 1667~MHz line because it places the strongest upper limit
on the OH column density among the 18~cm lines.  The continuum
flux density is interpolated at the expected frequency of the redshifted 1667~MHz line
from continuum measurements in the literature following the same method used for the 
21~cm spectral observations (Section~\ref{sec:flux}). When the 1667~MHz line could not be 
observed due to RFI, the 1612 or 1720~MHz transitions are used 
as the reference instead.

\subsection{Line Detections and Limits}
\begin{figure*}
\minipage{0.35\textwidth}
  \includegraphics[width=\linewidth]{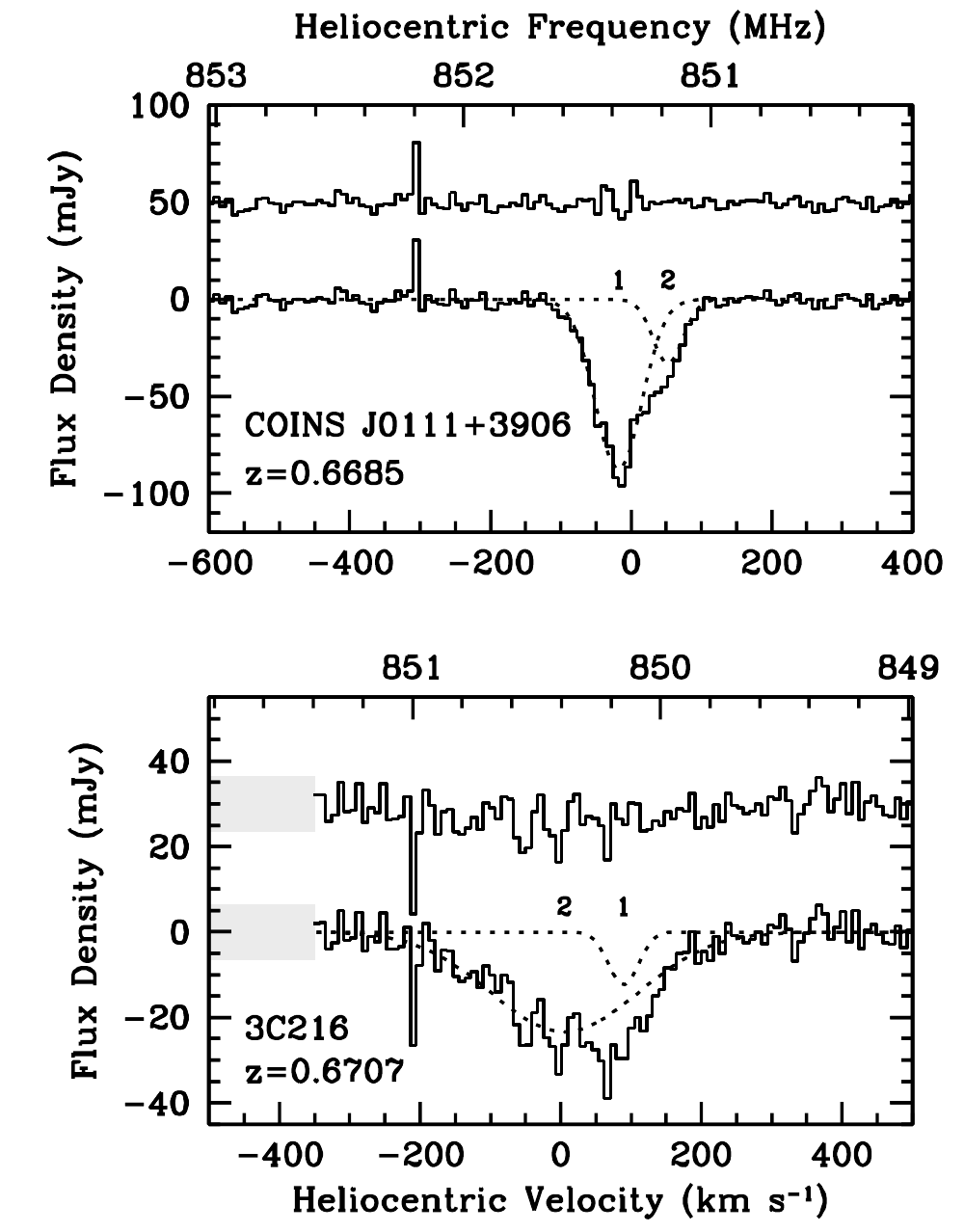}
\endminipage
\minipage{0.32\textwidth}
  \includegraphics[width=\linewidth]{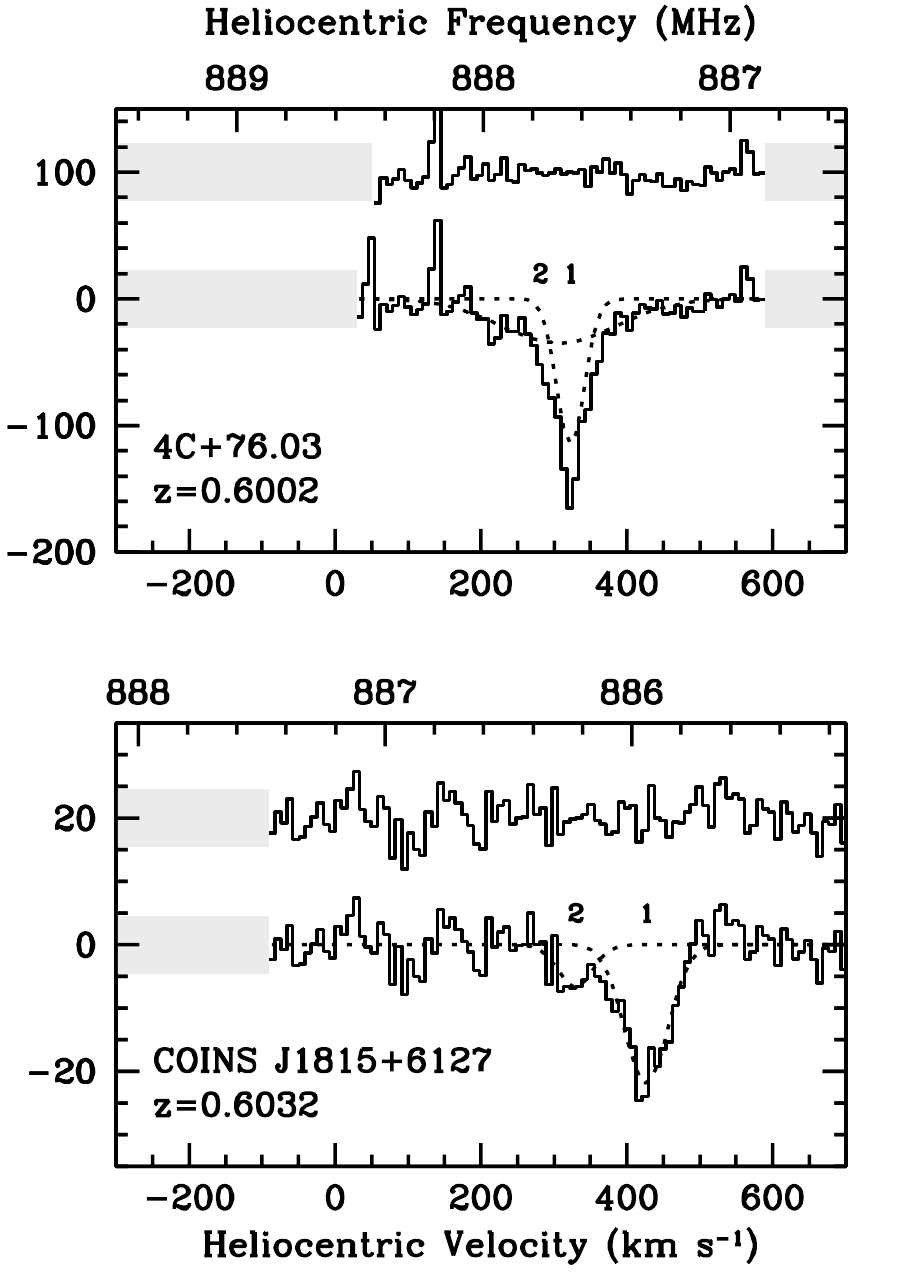}
\endminipage
\minipage{0.32\textwidth}
  \includegraphics[width=\linewidth]{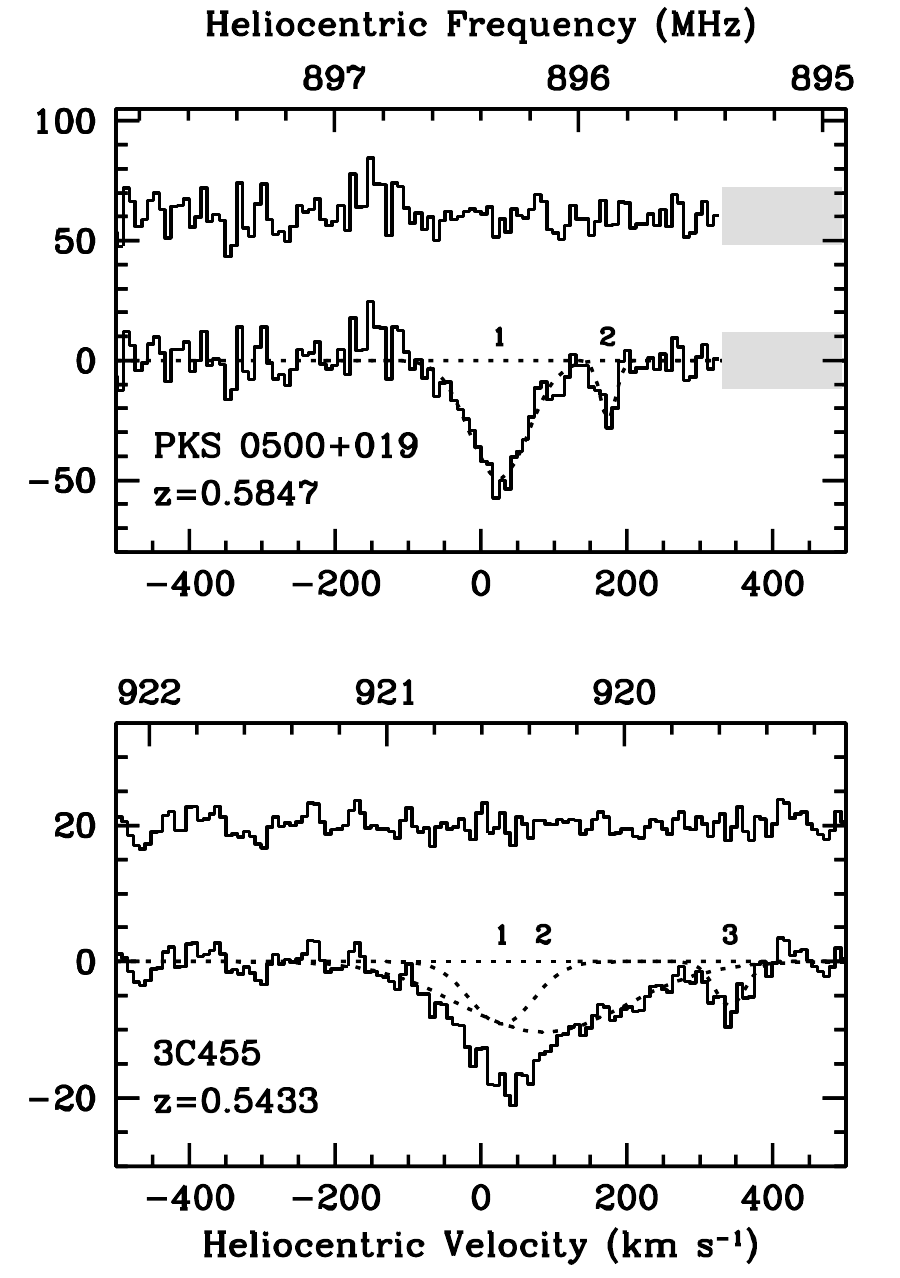}
\endminipage
  \caption{Intrinsic HI 21~cm absorption line systems. Dotted lines indicate
Gaussian fits, numbered by component listed in Table \ref{tab:detections}.
Upper spectra show the residual spectrum, offset for clarity. The velocity 
scale is in the rest frame of each object, defined by the optical heliocentric redshift
of each radio source (Table \ref{tab:table1}). Spectral regions lost to
radio frequency interference are greyed out.  %The listed redshift for each source is the redshift of the central absorption system.  
 }\label{fig:detected}
\end{figure*}

We detect intrinsic HI 21~cm absorption in 6 of the 108 RFI-free
compact radio sources in our sample. The detection 
rate is 5.6\%, substantially lower than 
in previous intrinsic HI absorption line surveys (Section~\ref{sec:discussion}).
All of these absorption lines are detected in previous surveys (see Table~\ref{tab:HI} for references).
The detected line spectra are shown in Figure~\ref{fig:detected},
where velocities are referenced to the optical heliocentric redshift. Table~\ref{tab:detections} lists the measured and derived properties of the detected HI 21~cm absorption lines.

Table~\ref{tab:HI} lists the measured upper limits to the column densities, calculated according to Section~\ref{sec:columnHI} and further discussed in light of previous studies in Section~\ref{sec:previous}. We show the spectra of the 102 RFI-free sources that show no significant HI 21~cm absorption lines and are predominately free from RFI in the Appendix. 

PKS~0742+10, at $z=2.64$, shows a redshifted negative spectral feature with a signal to noise of 3.4$\sigma$.  
We compare our spectrum to that of \citet{curran13}, also obtained at the GBT. The \citet{curran13} spectrum 
shows narrow RFI at the frequency of the potential absorption line. This suggests
that the feature in our spectrum is likely to be an incompletely flagged or low-level RFI feature.
We therefore treat PKS 0742+10 as a non-detection in all subsequent figures and statistics, with the rms noise measured away from the spectral feature.  

None of the four 18 cm OH lines are detected in absorption or emission 
in 102 RFI-free spectra out of the 125 objects observed.  
Table \ref{tab:OH} lists the properties of the OH sample, and 
Figure \ref{fig:OH_plot} in the Appendix shows the OH 1667 MHz  non-detection spectra of the four
21~cm absorbers in our sample with RFI-free OH spectral observations.

\subsection{Flux Densities}\label{sec:flux}
The nature of our observations with the GBT maximizes the on-time observation of targets and as a result does not give the necessary observations to flux calibrate the data and directly measure the continuum flux density. Moreover, for position-switched observations, RFI and instrumental effects frequently result in negative continuum values, artificial continuum slopes, and bandpass structure. 
The continuum is however paramount to obtain the optical depth and measure the column density or place a limit in the cases of non-detections (see eqn. \ref{eq:NHIdet}). We therefore obtain the continuum flux density at the frequency of the redshifted HI and OH
line of each source from an interpolation between
extant continuum measurements in the NASA Extragalactic Database (NED)
using a single spectral index power-law fit (a linear fit in
log~$S_{\nu}$--log~$\nu$ space). In order to cope with the heterogeneous
literature continuum measurements, we try whenever possible to select
values published from the same catalogs and pass bands, most often
observations at 1410, 1340, 750, 635, 408, and 365~MHz
\citep{laing80, large81, ficarra85, white92, douglas96, renge97, stanghellini98, condon98, stanghellini05, orienti07, petrov08}.
The uncertainties in the interpolated 
continuum flux densities listed in Tables \ref{tab:HI} and \ref{tab:OH} are obtained from estimates of the
1$\sigma$ confidence intervals of the power law continuum fits for each object. This treatment neglects
potentially significant time variation in radio source fluxes.

\subsection{Column Densities}\label{sec:column}
\subsubsection{HI Column Density}\label{sec:columnHI}
Column densities are derived from the
integrated optical depths of spectral lines. The column density associated with a given 21-cm line optical depth is calculated according to \citet{wolfe75} as
	\begin{equation}\label{eq:NHIdet}
N_{\rm HI} = (1.8 \times 10^{18}\ {\rm cm}^2) \ {\frac{T_s}{f}}  \int \tau \ dv,
	\end{equation}
where $T_s$ is the spin temperature of the 21~cm line in Kelvin, $f$ is the
fraction of the area of the continuum source covered by the absorber, and
$\int \tau \ dv$ is the optical depth $\tau$ integrated across the velocity 
span of the line, in \kms, calculated from the continuum fluxes acquired in Section~\ref{sec:flux} and the strength of the absorption line.  
A Gaussian absorption profile can be integrated:
$\int \tau \ dv = \sqrt{\pi/\ln 2}\ \tau_{\rm max}\,\Delta v/2$, where 
$\tau_{\rm max}$ is the peak optical depth of a line with a full width half max (FWHM) of $\Delta v$. The column densities of the six re-detections of 21 cm absorption are calculated according to Eq.~\ref{eq:NHIdet} and are listed in Table~\ref{tab:detections}. We use both Gaussian fits and direct integration for the total HI column density of the 21~cm detections.
 
For the non-detections, we assume optically thin lines, and the $3\sigma$ upper limit on the 
column density can be approximated by 
	\begin{equation}\label{eq:NHI}
N{\rm _{HI, 3\sigma}} < (1.9\times 10^{18} {\rm cm}^{-2})\ \frac{T_s}{f}\ \tau_{3\sigma} \ \Delta v\  ,
	\end{equation}
where $\tau_{3\sigma} \approx 3\sigma/S$, $\sigma$ is the rms noise, and 
$S$ is the continuum flux density of the radio source at the redshifted line frequency.

All HI column densities are computed assuming $T_s = 100$~K and a uniform covering factor of $f = 1$, resulting in an HI column density limit in the CNM phase of the ISM. We note that this is not the total HI column density as the 21~cm transition is not a reliable tracer of the gas in WNM due to the optical depth inversely related to the spin temperature. 

Non-detection column density limits assume $\Delta v=30$~\kms. Table~\ref{tab:HI} 
lists the column densities and limits for the non-detections and all upper limit spectra are shown in the Appendix.  
Figure~\ref{fig:NHIz} shows the HI column density distribution as a function of redshift
for the 108 RFI-free sources.

\subsubsection{OH Column Density}
\begin{figure}
\epsscale{1.2}
\plotone{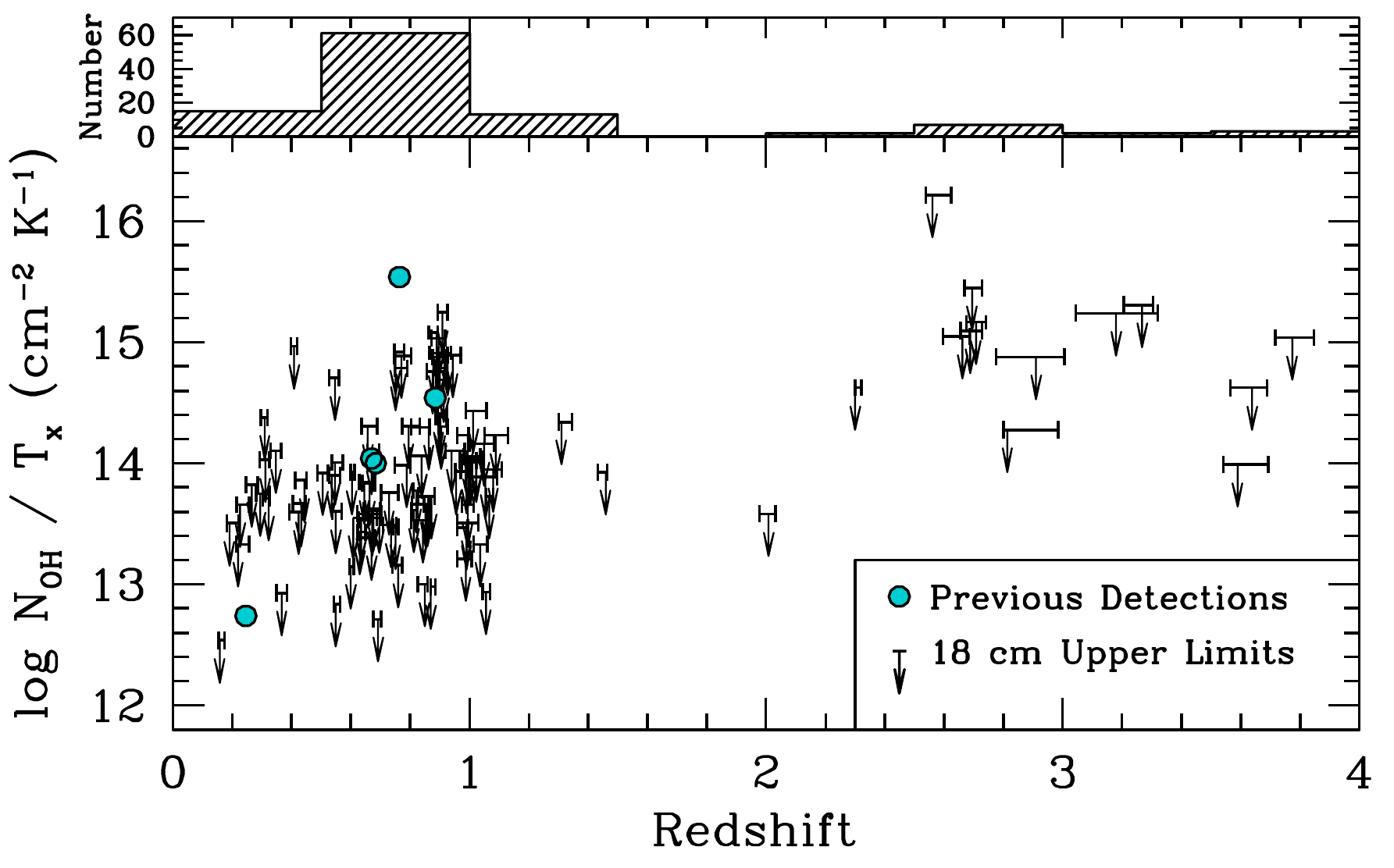}
\caption{The lower panel shows
  the 3$\sigma$ upper limit OH column density versus the optical redshift. The
  horizontal bar indicates the redshift search region, typically determined
  by RFI conditions, and the downward-pointing arrow is centered on the
  systemic redshift of each object. The upper panel indicates the number
  of objects surveyed in each $\Delta z=0.5$ redshift bin (black, hatched). We show previous OH 1665~MHz absorption detections as teal symbols.
\label{fig:NOHz}
}
\end{figure}

For the redshifted OH 18~cm line non-detections, we use
the 1667~MHz OH transition to place the strongest upper limit on the
OH column densities via
	\begin{equation}
N{\rm _{OH, 3\sigma}} < X \times 1.0645 \ \frac{T_x}{f} \ \tau_{3\sigma}\ \Delta v\ , 
	\end{equation}
where $X = 2.38 \times 10^{14}$~cm$^{-2}$ \citep{curran08} for the 1667~MHz transition,
and $T_x$ is the excitation temperature of the OH 1667 MHz line.
When the 1667~MHz line could not be observed due to RFI, the OH column
density limit is calculated using the 1612~MHz or 1720~MHz transition, which have
$X = 2.14 \times 10^{15}$~cm$^{-2}$ ($X_{1612~ \rm{MHz}} = 9 \times X_{1667~ \rm MHz}$, following the 
18 cm OH line ratios in thermodynamic equilibrium).  We make no assumption about the OH
line excitation temperatures and leave $T_x$ as a free parameter in our tabulation and analysis.  
As with the HI lines, for the OH lines we assume a uniform covering factor of $f = 1$, and a
line width of $\Delta v=30$~\kms.  
Figure \ref{fig:NOHz} shows the OH
column density limits as a function of redshift for the 102 RFI-free sources in the sample.

Because we find no OH detections, as well as no new HI detections, we are unable to place any meaningful constraint on possible redshift variations in the fundamental physical constants, discussed in Section~\ref{sec:intro}.

\subsection{Comparison to Previous Work}

\subsubsection{HI 21~cm Absorption Studies}\label{sec:previous}
This intrinsic HI 21~cm absorption search toward 144 compact radio
sources (Table~\ref{tab:table1}; one additional source is only observed for the four OH lines, for a total sample size of 145) obtained 108 spectra including 
six re-detections, with 36 sources remaining ambiguous due to RFI.  
21 objects in the survey sample are previously observed for intrinsic HI absorption
including nine known 21~cm line absorbers in the sample. 

We made no
new 21~cm absorption detections, however, we do re-detect six of the nine known HI absorption
systems (Table~\ref{tab:detections}). We are not able to re-detect the three
absorption line systems toward 3C49, B3~1355+441 \citep{ver03} or 3C190
\citep{ishwara03} due to RFI.

Our spectra of 3C138, 3C147, 0239+108, OE+131, TXS 0902+490, HB89~0954+658, 1004+141, FBQS~J1159+2914, 1418+546, 1642+690, 3C380, 4C+29.56, 2007+222, and
PKS~2149+056 improves the previous upper limit measurements \citep{ver03, carilli98, curran17, curran19, aditya18a, aditya18b}.
Three objects, B3~0248+430, SBS 0804+499, and 3C147, were impacted by RFI in our observations, preventing 
independent upper limits on previous measurements.  
Five sources, 3C138, HB89 0754+100, COINSJ~1546+0026, PKS 2121$-$01, and HB89~2342+821 have
column density limits comparable to previous surveys. Only two limits on our sources, B2~0738+31 and COINS
J2022+6136, did not improve on previously published column density limits. 
The remaining 79 sources have (to our knowledge) never been searched for intrinsic HI absorption
before. Table~\ref{tab:HI} lists the column densities or column density limits from our survey and
previous surveys along with references, if applicable. All non-detections quoted at the $3\sigma$ level use the assumptions listed in Section~\ref{sec:column}. 

The six detected absorbers all have column densities near or greater than the DLA limit of N$_{\rm HI}>2\times10^{20}$~cm$^{-2}$ and we reached adequate sensitivity to have detected most known redshifted 21~cm absorbers if they had been present in our sample. 70\% of our non-detections have 3$\sigma$ upper limits below the DLA column density threshold for the detected absorbers. This suggests that while we have the sensitivity to detect new systems of cold gas (T$\sim$100~K) at sub-DLA column densities, there must be physical reasons for the dearth of absorption lines in these redshifts radio sources in this survey, which we investigate further in Section~\ref{sec:UV}.

\subsubsection{OH 18~cm Absorption Studies}
We searched 125 radio sources for the four OH 18~cm
lines, which can appear in either emission or absorption.
The spectra of 102 objects are unaffected by RFI for at least one of the four OH
transitions.  
No OH emission or absorption lines are detected in any of the RFI-free spectra and the OH column density upper limits are reported in Table~\ref{tab:OH}. Only one source in our sample, B3~1355+441, has prior OH observations \citep{curran06}. We report a 3$\sigma$ upper limit on the OH column density of $N_{\rm OH}/T_x <2.1\times10^{13}$~cm$^{-2}$~K$^{-1}$, an improvement over of the 3$\sigma$ measurement of $N_{\rm OH}/T_x < 11\times10^{13}$~cm$^{-2}$~K$^{-1}$ from \citet{curran06}.

Four RFI-free OH observations toward known 21~cm absorbers (Figure~\ref{fig:OH_plot}) do not detect any OH lines despite reaching very low optical depths, 
$0.002 < \tau_{3\sigma} < 0.01$ (Table~\ref{tab:OH}).  
OH column density upper limits for the full sample span the range 
$10^{12}$~cm$^{-2}$~K$^{-1}<N_{\rm OH}/T_x < 10^{16}$~cm$^{-2}$~K$^{-1}$, 
consistent with upper limits quoted
in previous surveys for non-detections of OH 18~cm absorption \citep{gupta06, curran08, curran11}.

\section{Analysis}\label{sec:analysis}
We detect no new 21~cm absorption systems in our survey. We do, however, re-detect six known absorbers, for a detection rate of 5.6\%. The following sections discuss factors that may negatively impact the ability to detect neutral HI gas, especially at high-redshift.

\subsection{Radio Continuum Luminosity}
One possible explanation of our low detection rate, especially at
high redshift, is a high spin temperature. As shown in Equation~\ref{eq:NHI},
the measured HI column density is proportional to the (unknown) spin temperature $T_s$ of
the absorbing gas.  
Previous studies investigated the impact of the local radio radiation field in driving the spin temperature up in associated absorption systems \citep[e.g.,][]{gupta06, cur10}, finding little to no significance dependence of the radio luminosity on the detection rate or $N_{\rm HI}$. If the spin temperature is coupled to the local radio radiation temperature, an increase would cause weaker line absorption for a given column density. However, recent work by \citet{aditya18a, aditya18b} suggest that gas excitation from the radio continuum may be responsible for the dearth of 21-cm absorbers detected at high redshift. This suggests that the spin temperature of the gas, a measure of the excitation from the lower hyperfine level \citep{field59}, can be raised by excitation to the upper hyperfine level by rest-frame 1420~MHz photons. \citet{curran19} has shown that this is unlikely and that the photoionization from the UV luminosity is dominant factor over the radio luminosity in negatively impacting the detection rate of 21~cm absorbers (see Section~\ref{sec:UV}). 

We test this possibility of the impact of the incident 21~cm luminosity on the absorption detection rate by 
calculating the specific radio continuum luminosity in the vicinity of the 21~cm line in the rest frame of
each host galaxy \citep{vang89}:
	\begin{equation}
L_{\nu} = \frac{4\pi  D_{\rm L}^2 \ S_{\nu} }{1+z}~\rm{W~Hz}^{-1}, 
	\end{equation}
where $D_{\rm L}$ is the luminosity distance of the host galaxy, $S_{\nu}$ is the observed
21~cm flux density at the frequency of the redshifted 21~cm observations,
as calculated in Section \ref{sec:results} and listed in Table \ref{tab:HI}, and
$z$ is the redshift.  

Table \ref{tab:magnitudes} lists the radio continuum luminosities for the sample.
The median radio continuum luminosity of the detections is $\log (L_{\rm
  Radio}/{\rm W~Hz}^{-1}) = 27.5\pm0.4$, consistent with the median luminosity for the
non-detections at $27.5\pm0.7$.  
A Kolmogorov-Smirnov Test (KS-test) shows that the 21~cm rest-frame luminosity differs between the detections and non-detections at less than 1$\sigma$ confidence level; the 21~cm luminosity does not appear to be a major factor in 21~cm absorption and cannot explain our low detection rate, also found in \citet{cur10}.  

\begin{figure}
\epsscale{1.2}
\plotone{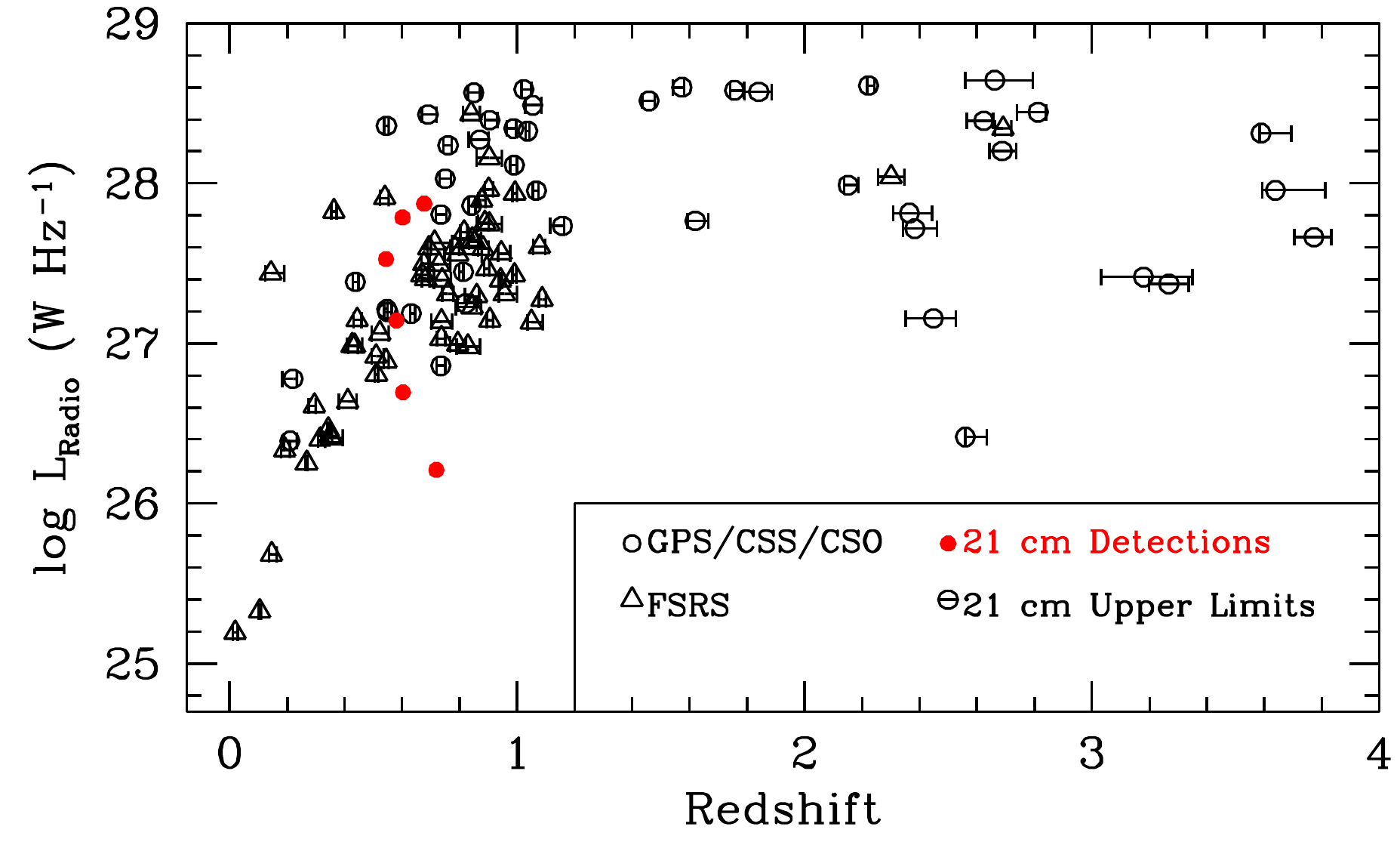}
\caption{Radio luminosity versus redshift for RFI-free sources. We use
  the optical redshift for 
  sources not detected in 21~cm and the absorption redshift for the six
  21~cm detections. The horizontal bar on the 21~cm non-detections
  indicates the redshift search 
  region for each source, generally determined by the RFI conditions. The
  solid red symbols represent 21~cm detections and unfilled black symbols
  represent the non-detections. The shape of the symbols represent the
  radio source identification: circles represent GPS, CSS, and CSO
  sources, and triangles represent FSRSs.  
\label{fig:Radz}
}
\end{figure}

\begin{figure}
\epsscale{1.2}
\plotone{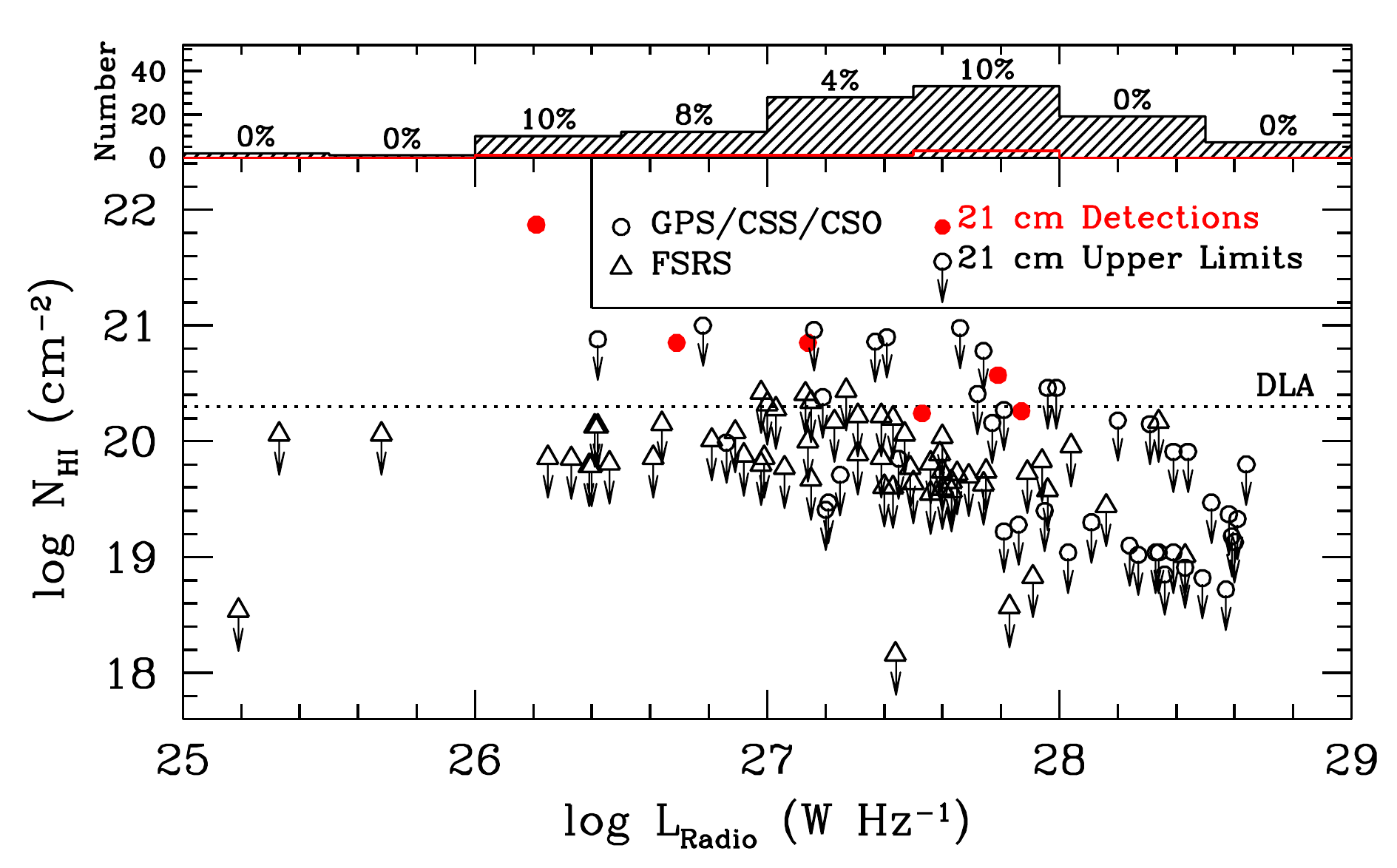}
\caption{The lower panel shows
  the HI column density (upper limit in the case of non-detections) as a
  function of the specific radio luminosity. The horizontal dotted line indicates
  the canonical threshold for DLA systems. The solid red symbols represent 21~cm
  detections and unfilled symbols represent the non-detections. The
  shape of the symbols represent the radio source identification: 
  circles represent GPS, CSS, and CSO sources, and triangles
  represent FSRSs. The upper panel indicates the total number of
  objects surveyed in each $\Delta L_{\rm Radio}=0.5$ dex bin (black,
  hatched histogram), the number of intrinsic HI 21~cm absorption lines
  detected (red histogram), and the 21~cm absorption fraction (text
  percentages above each bin).   
\label{fig:NHI_RL}
}
\end{figure}

With the exception of a few low-luminosity radio sources, the radio
luminosity-redshift distribution shows the Malmquist bias
for flux-limited surveys: the most luminous objects lie at the 
largest distances (Figure \ref{fig:Radz}). 
As expected, the detections show an inverse correlation of radio luminosity with column density (Figure \ref{fig:NHI_RL}). This, however, is an observational bias because the highest flux
density sources typically have the highest luminosities (Figure \ref{fig:Radz}), and
high flux densities provide greater column density sensitivity.

\subsection{Ultraviolet Luminosity}\label{sec:UV}
Another possible explanation for this survey's lack of intrinsic 21~cm absorption at high redshift
is the hypothesis proposed by \citet{curran08} that AGN with a high ultraviolet ($\lambda_{\rm UV} = 1216$~\AA) luminosity ($L_{\rm  UV}>10^{23}$~W~Hz$^{-1}$) may inhibit 21~cm absorption. If the UV flux of the central AGN is sufficient to ionize (or simply heat) the neutral gas within the host galaxy, UV-bright galaxies will be biased against the detection of 21 cm absorbers. These UV bright sources are also likely to be unextincted and may simply have less gas and dust along the line of sight compared to extincted sources, lowering the ability to detect absorption systems as sensitivity depends on both the covering fraction of the gas against the background source and the spin temperature of the gas. The UV luminosity bias will be compounded with increasing redshift because flux-limited surveys will naturally select UV-luminous objects at all redshifts.

We have searched the literature for optical and near-IR photometry of each
source in our sample to estimate the rest-frame UV luminosity following the method in \cite{curran08}.  
Table \ref{tab:magnitudes} lists the optical and near-IR magnitudes for the sources searched for 21~cm absorption.
Of our detected absorbers, 4C+76.03 does not have adequate available optical or near-IR observations. For this source, along with six sources with non-detections, we are unable to estimate the UV-optical slope in order to measure the expected UV luminosity. We therefore do not calculate a UV luminosity for these sources and they have been excluded from all plots and analysis related to the UV luminosity. In conclusion, we have five detections in 21~cm absorption and 96 sources with upper limits that we are able to calculate a UV luminosity for in order to test the hypothesis proposed by \citet{curran08} for the ionization of the atomic gas in UV-bright radio sources. 

Using visible band
magnitudes and attempting to maintain consistency in the heterogeneous 
sample and sources of magnitudes, the extrapolation or  
interpolation of the rest-frame UV luminosity at 1216~\AA\ is 
performed using a power-law fit to the $B$ and $R$ bands as these two bands are readily available for most for the sample.  
For sources where either of those bands are not available, observations at $V$ or $K$ bands are 
substituted in that order. We require a minimum of two bands to perform the power-law fit, for galaxies with more than two available magnitudes, we use all available optical and near-IR bands. Objects with Sloan Digital Sky Survey
(SDSS) $u'g'r'i'z'$ magnitudes are converted to Johnson 
magnitudes using the methods described in \citet{smith02} and are not corrected for intrinsic source extinction.
  
The fitting is done by converting the visible
band magnitudes to fluxes using Vega as a zero-point reference, with values
for Vega in each band taken from \citet{bessell79}, \citet{campins85}, and
\citet{schneider83}. Under the assumption that a single power law applies from the observed optical wavelengths down to 1216~\AA, the UV flux density in the rest-frame of the host galaxy is obtained from the 
observed flux $F_1$ at wavelength $\lambda_1$ from \citet{curran08} as,
	\begin{equation}\label{eq:UVflux}
F_{\lambda_{\rm UV}(1+z)} = F_1 \, \left[ \frac{\lambda_{\rm UV} \
    (1+z)}{\lambda_{1}}\right]^{\alpha}, 
	\end{equation}
where the spectral index $\alpha$ is estimated as $\alpha =
\log(F_2/F_1)/\log(\lambda_2/\lambda_1)$ for two observations at
bands $\lambda_1$ and $\lambda_2$, with fluxes $F_1$ and $F_2$.  This allows us to extrapolate from the observed visible-band photometry to obtain a UV flux at the observed wavelength $\lambda_{\rm UV}(1+z)$. 
The specific UV luminosity in the rest-frame of the host galaxy is then calculated as,
	\begin{equation} 
L_{\lambda_{\rm UV}} = \frac{4\pi D_{\rm L}^2 \ F_{\lambda_{\rm
      UV}(1+z)}}{1+z}~\rm{W~Hz}^{-1}, 
	\end{equation}
where $D_{\rm L}$ is the luminosity distance to the host galaxy and $F_{\lambda_{\rm UV}(1+z)}$ is the rest-frame UV flux calculated from Eq. \ref{eq:UVflux}. 

%LL_pvalue_stats.py
The median and 1$\sigma$ values of the UV luminosity of the full sample of 101 sources (5 detections, 96 non-detections, 7 sources with no L$_{\rm UV}$ measurement)
is $\log(L_{\rm UV}/{\rm W~Hz}^{-1}) = 23.1 \pm 1.4$, consistent with the UV luminosity threshold of 10$^{23}$~W~Hz$^{-1}$. The median luminosity of the HI absorption detections is 
$\log (L_{\rm UV}/{\rm W~Hz}^{-1}) = 20.5\pm1.1$, and the median luminosity of the 
non-detections is $\log (L_{\rm UV}/{\rm W~Hz}^{-1}) = 23.2\pm1.4$. If we only consider systems below the $L_{\rm UV} < 10^{23}$~W~Hz$^{-1}$ threshold, this reduces the number of RFI-free sources searched for 21~cm absorption from 101 to 47 targets. The detection rate for this subset of UV-faint sources increases to 9.8\%. 

\begin{figure}
\epsscale{1.2}
\plotone{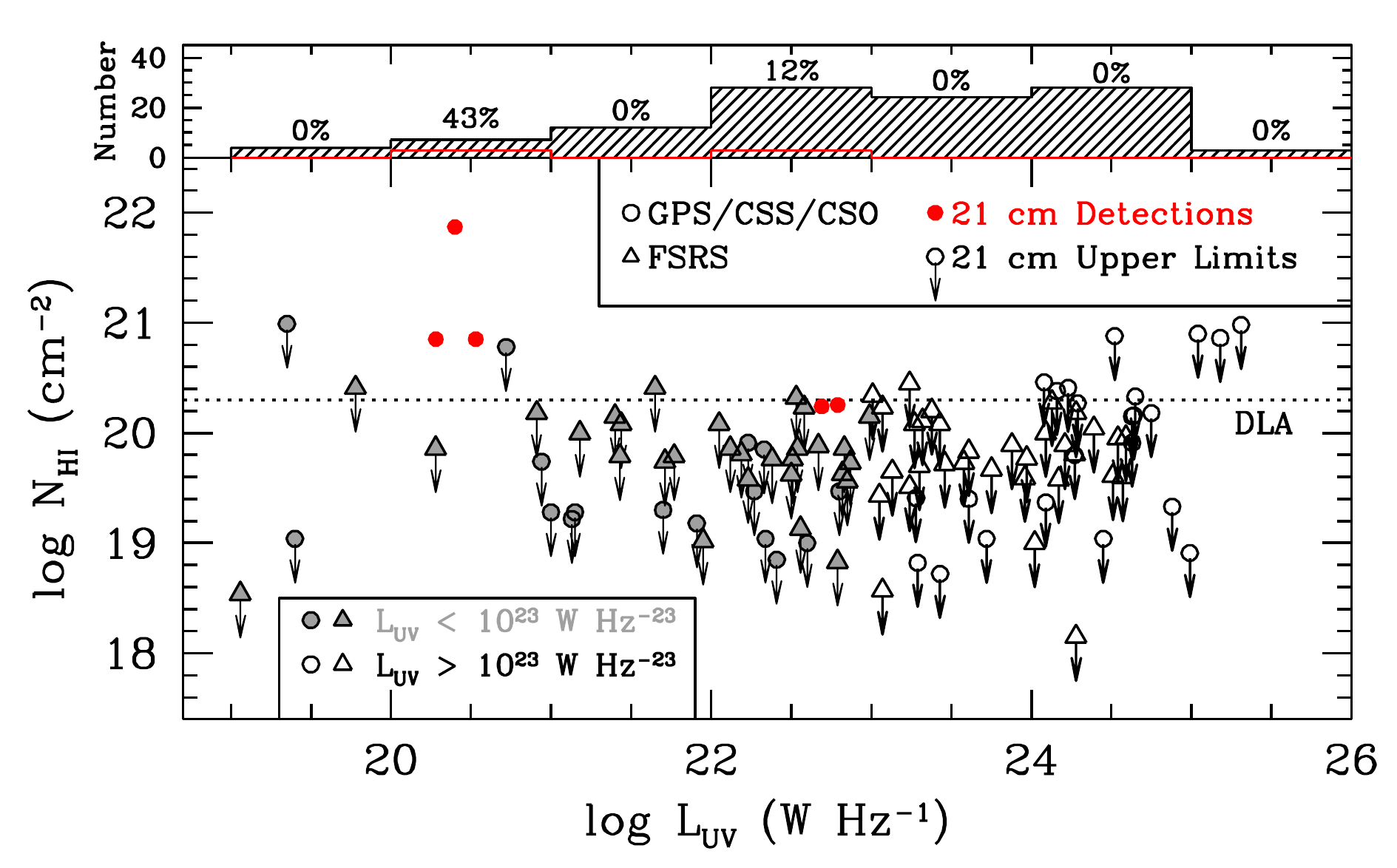}
\caption{HI column density N$_{\rm HI}$ (upper limit for sources not detected in 21~cm
  absorption) as a function of the ultraviolet luminosity (Section~\ref{sec:UV}). The horizontal dotted line
  indicates the threshold for DLAs. The solid red symbols represent 21~cm
  detections and downward pointing arrows are the upper limits for non-detections. Filled gray symbols represent the non-detections that are below the UV threshold of $10^{23}$~W~Hz$^{-1}$ and open symbols represent the non-detections that are above the UV threshold.
  The shape of the symbols represent the radio source identification:
  circles represent GPS, CSS and CSO sources and triangles represent
  FSRSs. The upper panel indicates the
  total number of objects surveyed in each $\Delta L_{\rm UV}=1$ dex 
  luminosity bin (black, hatched histogram), the number of intrinsic HI
  21~cm absorption lines detected (red histogram), and the 21~cm
  absorption fraction (text percentages above each bin). We confirm the findings of \citet{curran08} that 21~cm absorption is not detected in sources with a UV luminosity above the threshold of $10^{23}$~W~Hz$^{-1}$. However, we detect no new absorbers in the sample below this UV luminosity threshold.
\label{fig:NHI_L}
}
\end{figure}

\begin{figure}
\epsscale{1.2}
\plotone{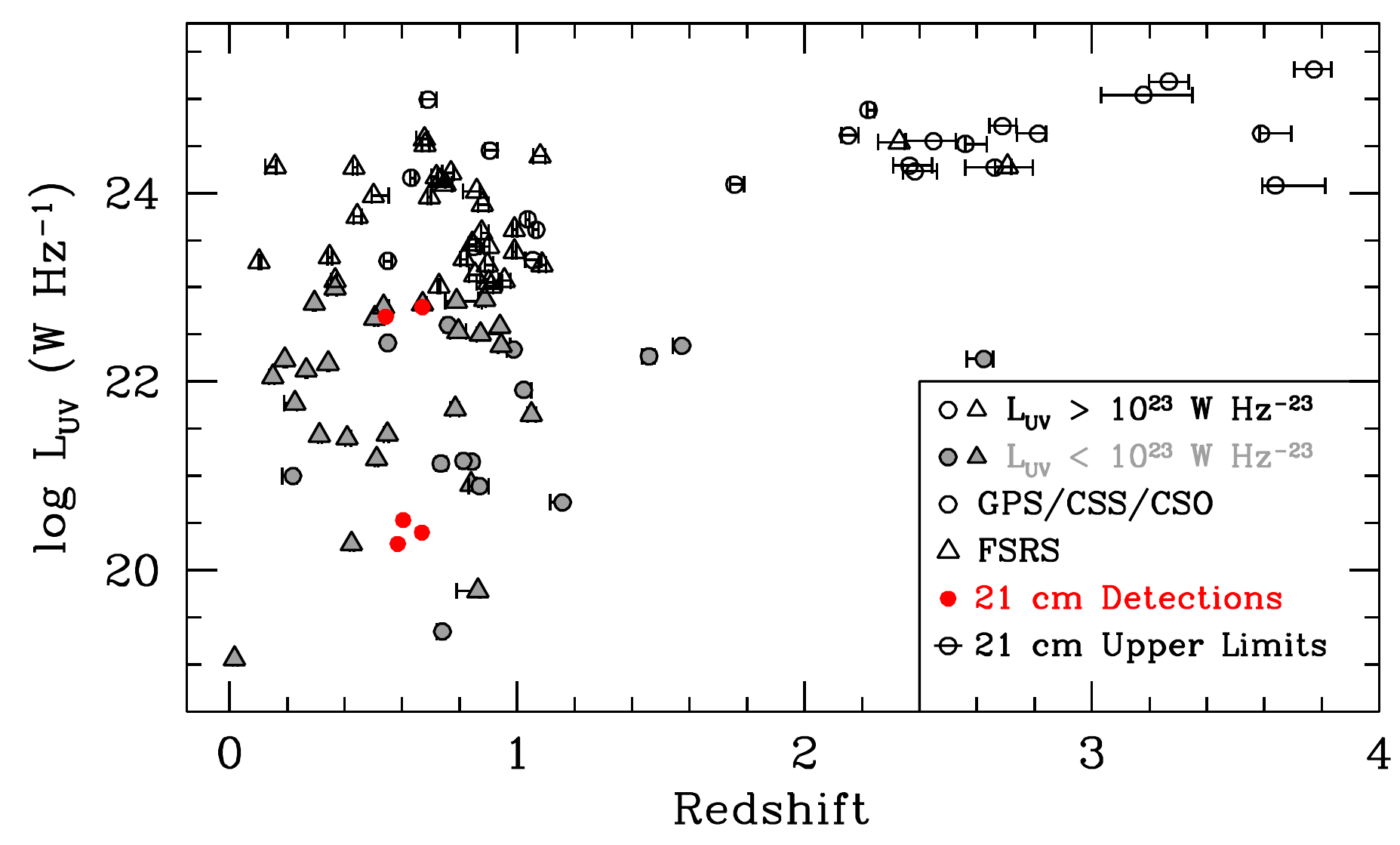}
\caption{UV luminosity versus redshift for our RFI-free sources.  We use
  the optical redshift for 
  sources not detected in 21~cm and the absorption redshift for the five
  21~cm detections with a measured UV luminosity. The horizontal bar on the 21~cm non-detections
  indicates the redshift search 
  region for each source, generally determined by the RFI conditions. The
  solid red symbols represent 21~cm detections and unfilled black symbols
  represent the non-detections. The shape of the symbols represent the
  radio source identification: circles represent GPS, CSS, and CSO
  sources and triangles represent FSRSs. 21~cm absorption systems have never been detected in sources with a UV luminosity above a threshold of $10^{23}$~W~Hz$^{-1}$. This figure demonstrates the difficulty in selecting sources below the UV luminosity threshold (filled gray symbols) at redshifts above $z\sim1$ that may possibly host 21~cm absorbers.
\label{fig:LUVz}
}
\end{figure}

Figure~\ref{fig:NHI_L} shows the distribution of the HI column density as a function UV luminosity. There are no HI 21~cm line absorbers associated with sources with UV luminosities above $\log (L_{\rm UV}/{\rm W~Hz}^{-1}) > 23$, further supporting the critical UV threshold above which 21~cm absorbers cannot be detected \citep{cur10, curran17}.  Figure~\ref{fig:LUVz} shows the UV luminosity as a function of redshift, demonstrating how the radio sources in this survey above a redshift of $z\sim1$ predominately have UV luminosities that are above the threshold. The requirement of UV/optical spectroscopic redshifts for radio spectral line surveys typically cause the high redshift sources to be the most luminous and are not ideal targets to show 21~cm or molecular absorption. %This highlights the effort that should be made to design surveys that target radio loud, UV faint sources that are more likely to host large reservoirs of cold gas detectable with the 21~cm transition. 

A KS-test comparing UV luminosities of the 21~cm detections to the non-detections, however, is not statistically significant, showing a 2.2$\sigma$ difference between the samples. While the detections and non-detections do not appear to be drawn from distinct populations in UV luminosity, our sample of detections may simply be too small for statistical significance. 

\begin{figure}
\epsscale{1.2}
\plotone{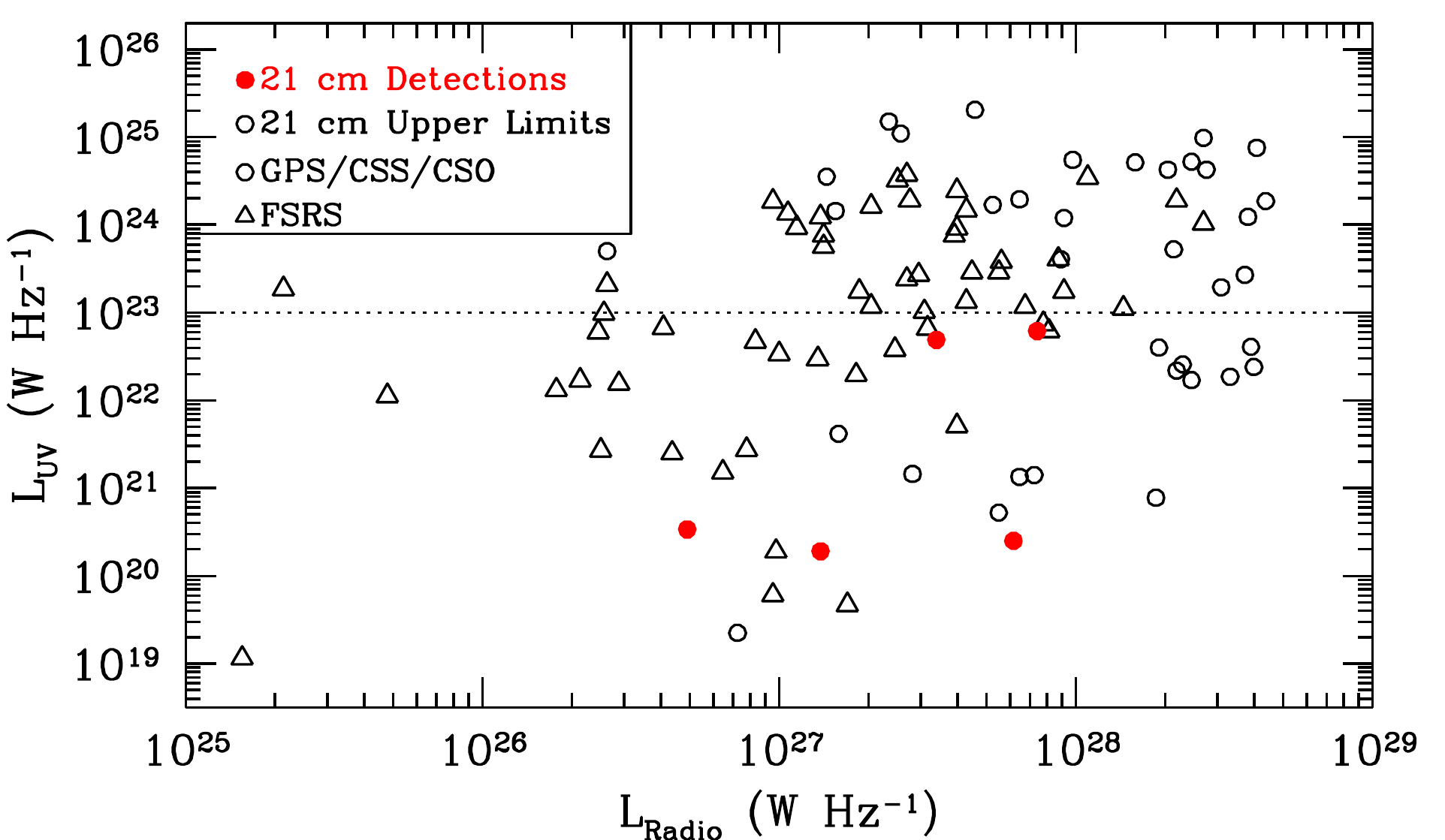}
\caption{Ultraviolet luminosity versus radio continuum luminosity at the frequency of the redshifted 21~cm line. The solid red symbols represent 21~cm detections and unfilled
  black symbols represent non-detections. The shape of the symbols represent the radio source
  identification: circles represent GPS, CSS, and CSO sources and 
  triangles represent FSRSs. The horizontal dotted line indicates the
  UV luminosity threshold ($L_{\rm UV} = 10^{23}$~W~Hz$^{-1}$) above which 21~cm absorption has not been detected. 
\label{fig:UVRad}}
\end{figure}

\begin{figure}
\epsscale{1.2}
\plotone{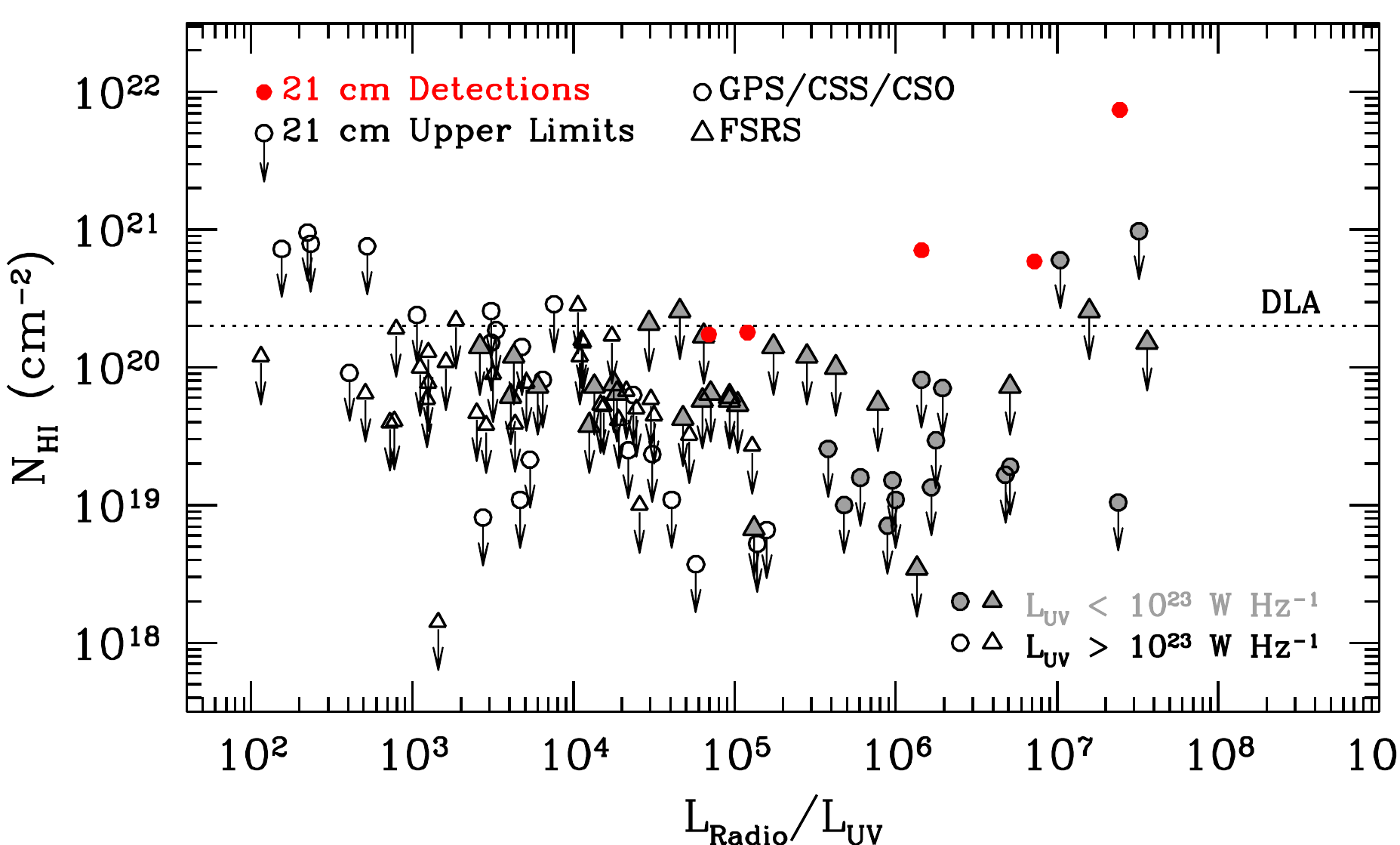}
\caption{HI column density (upper limit for sources not detected in 21~cm
  absorption) 
  versus the ratio of the radio to the UV luminosity (or flux density). The filled gray points
  represent HI 21~cm non-detections below the intrinsic UV luminosity threshold of
  $10^{23}$~W~Hz$^{-1}$ and black open symbols represent sources with an UV
  luminosity above the threshold. The solid red
  symbols represent 21~cm detections. The shape of
  the symbols represent the radio source identification: circles
  represent GPS, CSS, and CSO sources and triangles represent FSRSs.
  The horizontal dotted line indicates the threshold for damped Ly$\alpha$
  systems  
\label{fig:ratioL}}
\end{figure}

Figure \ref{fig:UVRad} shows the radio versus UV luminosities of the 
sample with a Pearson correlation coefficient of $r=0.35$, significant at the 3.7$\sigma$ level.  
As with all luminosity-luminosity plots of flux-limited samples, some degree of correlation 
can arise from the correlation of both luminosities with distance.
Figure \ref{fig:ratioL} shows the measured column density (or 3$\sigma$ upper limits) versus
the ratio of the radio-to-UV luminosity.  
The five HI 21~cm absorbers have a weak preference be located in objects that have a higher than median ratio, indicating their preference to reside in sources with low UV luminosity and/or higher radio luminosity compared to the non-detected sources. A KS test results in a D-value of 0.677, significant at 3.2$\sigma$. Future work with a larger number of detections will be able to further investigate the effect of the UV and radio luminosities on cold gas, especially investigating preference for 21~cm absorption in UV-faint, radio-loud sources.

\subsection{Physical Size}\label{sec:size}
The size of the radio source may also influence the detection of
HI 21~cm absorption.  If the radio lobes predominately
reside outside of the host galaxy, then the HI covering fraction 
will be small, and the apparent 21 cm optical depth will be correspondingly 
diminished.

The sizes of the radio sources in our survey are compiled from existing literature (Table~\ref{tab:HI}). Physical and projected sizes are calculated using the angular size distance for each redshift. Projection (and presumably relativistic beaming) can 
confuse analysis and interpretation because not all sources with small projected linear sizes are physically small and vice versa \citep[see, e.g.,][]{fanti89}.

Figure \ref{fig:NHIS} shows the HI column densities (upper limits
in the case of non-detections) as a function of the projected linear size of the
radio source. 
The detection of absorption systems does not seem to depend on source size and we find no correlation between the column density and the size of the source (Pearson correlation coefficient of $-$0.31). 
The detection rates among
the larger sources ($> 1$~kpc; 2 out of 21 sources) and the smaller sources
($< 1$~kpc; 4 out of 87 sources) are consistent with each other. 

\begin{figure}
\epsscale{1.2}
\plotone{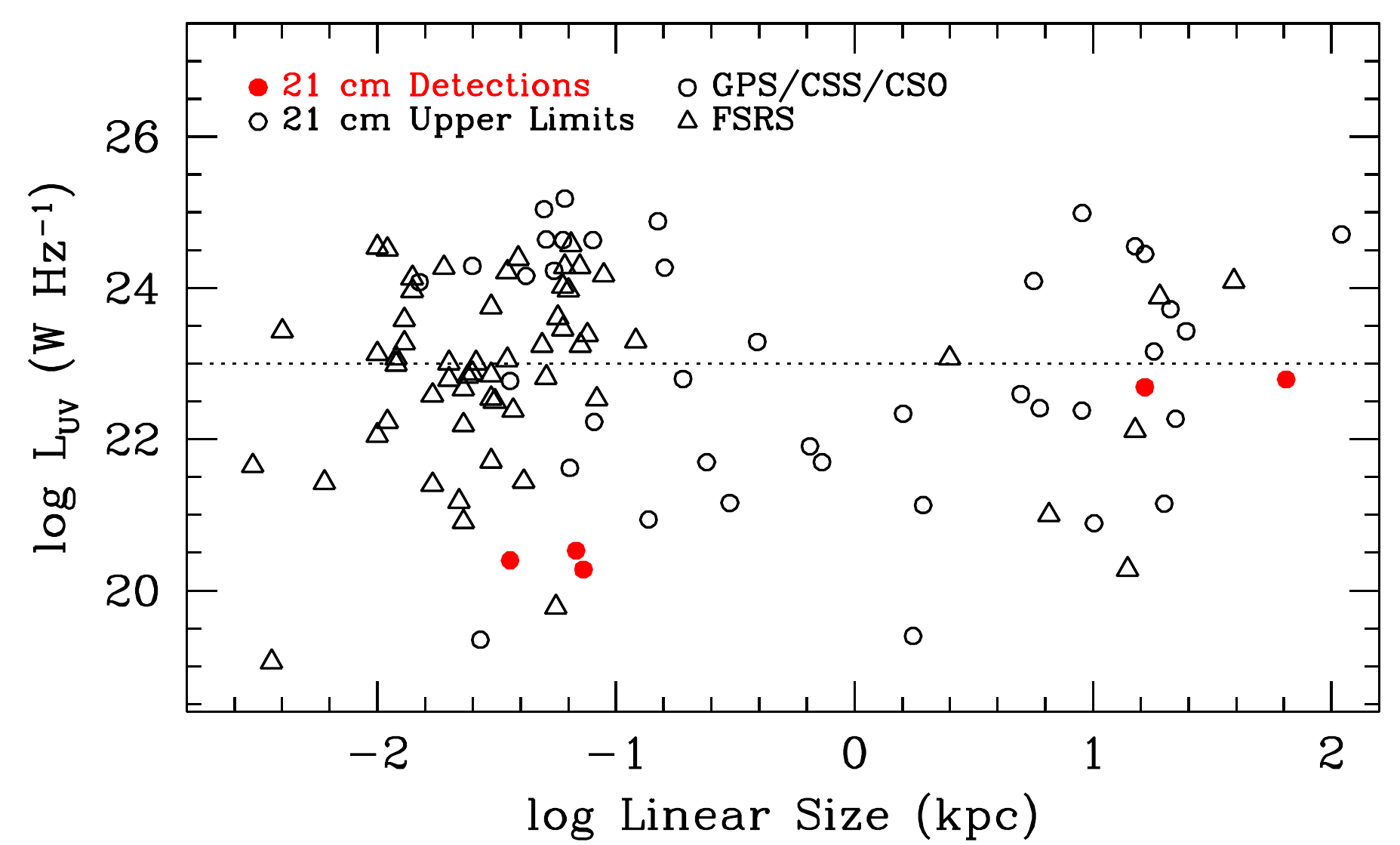}
\caption{Ultraviolet luminosity versus linear size of the radio sources. The
  solid red symbols represent 21~cm detections and black symbols
  represent the upper limits to the column density for non-detections. The
  shape of the symbols represent the radio source identification:
  circles representing GPS, CSS, and CSO sources and triangles
  represent FSRSs. The horizontal dotted line indicates the UV
  luminosity threshold of ($L_{\rm UV} > 10^{23}$~W~Hz$^{-1}$).   
\label{fig:UVS}}
\end{figure}

Figure \ref{fig:UVS} shows that UV luminosity does not correlate with source size
(correlation coefficient of $r=0.14$, non-significant at 1.4$\sigma$). Among the detections, the larger sources do exhibit a higher UV luminosity ($L_{\rm UV}>10^{22.5}$~W~Hz$^{-1}$) than the smaller sources
($L_{\rm UV}<10^{21}$~W~Hz$^{-1}$), but the correlation is not statistically significant for a sample of five detections. Among non-detections, smaller sources ($\leq1$~kpc) show a median UV luminosity of $\log L_{\rm UV} = 23.0\pm1.3$~W~Hz$^{-1}$, consistent with that of the larger sources ($> 1$~kpc) with a median of  $\log L_{\rm UV} = 22.6\pm1.5$~W~Hz$^{-1}$. The median UV luminosities likewise do not differ among radio source classification types: the median UV luminosity of the GPS, CSS, and CSO sources is $\log L_{\rm UV}=  22.8\pm1.6$~W~Hz$^{-1}$, while for FSRSs, the median value is $\log L_{\rm UV} = 23.1\pm1.2$~W~Hz$^{-1}$. 

\begin{figure}
\epsscale{1.2}
\plotone{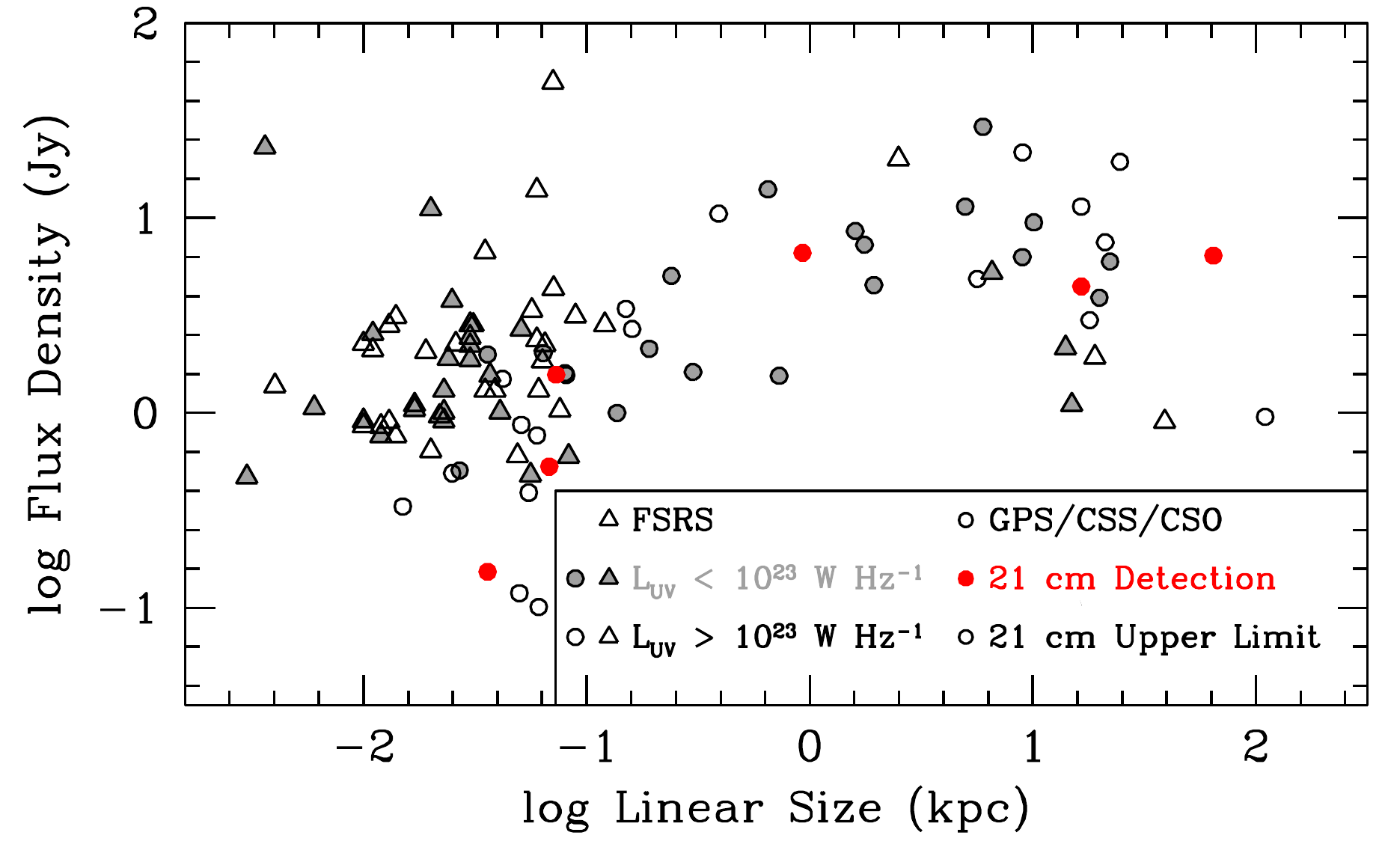}
\caption{Continuum flux density (Jy) as a function of the linear
  size of the radio emission structures in the sample. The
  solid red symbols represent 21~cm detections and unfilled black symbols
  represent non-detections. The
  shape of the symbols indicates the radio source identification: circles represent GPS, CSS, and CSO sources and triangles represent FSRSs.  
\label{fig:CntmSize}
}
\end{figure}

Figure \ref{fig:CntmSize} shows the relation between the radio continuum flux density and 
linear size of the radio jets, where the smallest radio sources tend to also be the faintest. These properties are countervailing in terms of 
the HI 21~cm absorption line detectability. Column density sensitivity
increases toward brighter objects, which tend to occur in the largest radio sources.  
However, the largest radio sources are the most evolved (thus possibly residing in galaxies with less
neutral gas) and most likely have the smallest covering fraction if the radio lobes are located outside the galaxy, reducing sensitivity for systems with lower column densities. We do not find that smaller sources have a higher rate of detection in 21~cm absorption, instead, we find that the detected absorbers cover the full range of jet sizes available in the survey, concluding that the size of the radio source is not a main contributor against the detection of HI 21~cm absorption.

\section{Discussion}\label{sec:discussion}
Some of the first surveys toward compact radio sources revealed intrinsic HI 21~cm detection rates of 30-60\% in low redshift radio galaxies \citep{morganti01, pih03, ver03, gupta06}. 
These previous surveys measure typical optical depth limits of $\tau_{3\sigma} \lesssim 0.001 - 0.2$ for non-detections. We recover a mean optical depth limit of $\tau_{3\sigma} < 0.023$ for our non-detections with a broad range from $10^{-6} < \tau_{3\sigma} < 0.15$, where 70\% of the non-detections lie below the DLA column density threshold. For sources in our survey that were also present in previous studies, we typically achieve greater sensitivity and are able to improve prior upper limits. 

The high success rate of 21~cm absorbers in these low-z studies have triggered follow up surveys to target compact radio sources in attempt to increase the detection rate of 21~cm absorption systems. The high success rates of these low-redshift surveys are in stark contrast, however, with surveys aimed toward high redshift $z\gtrsim1$ targets, where the detection rate plummets to single digits \citep{curran08, aditya16, curran17}, hinting at a connection between the redshift of the source and the success of detecting 21~cm absorption lines. 
While our sources span the redshift range $0.2<z<3.8$, 75\% of our RFI-free sample lie at redshifts $z<1$ and we recover a detection rate of less than 6\%, with all absorbers identified in prior surveys. Lack of new detections, especially at low redshift, while reaching the sensitivity to re-detect known absorbers, demonstrates that there have to be external physical factors at play that negatively impact the detection rate of neutral gas absorbers in our sample.

\begin{figure}
\epsscale{1.2}
\plotone{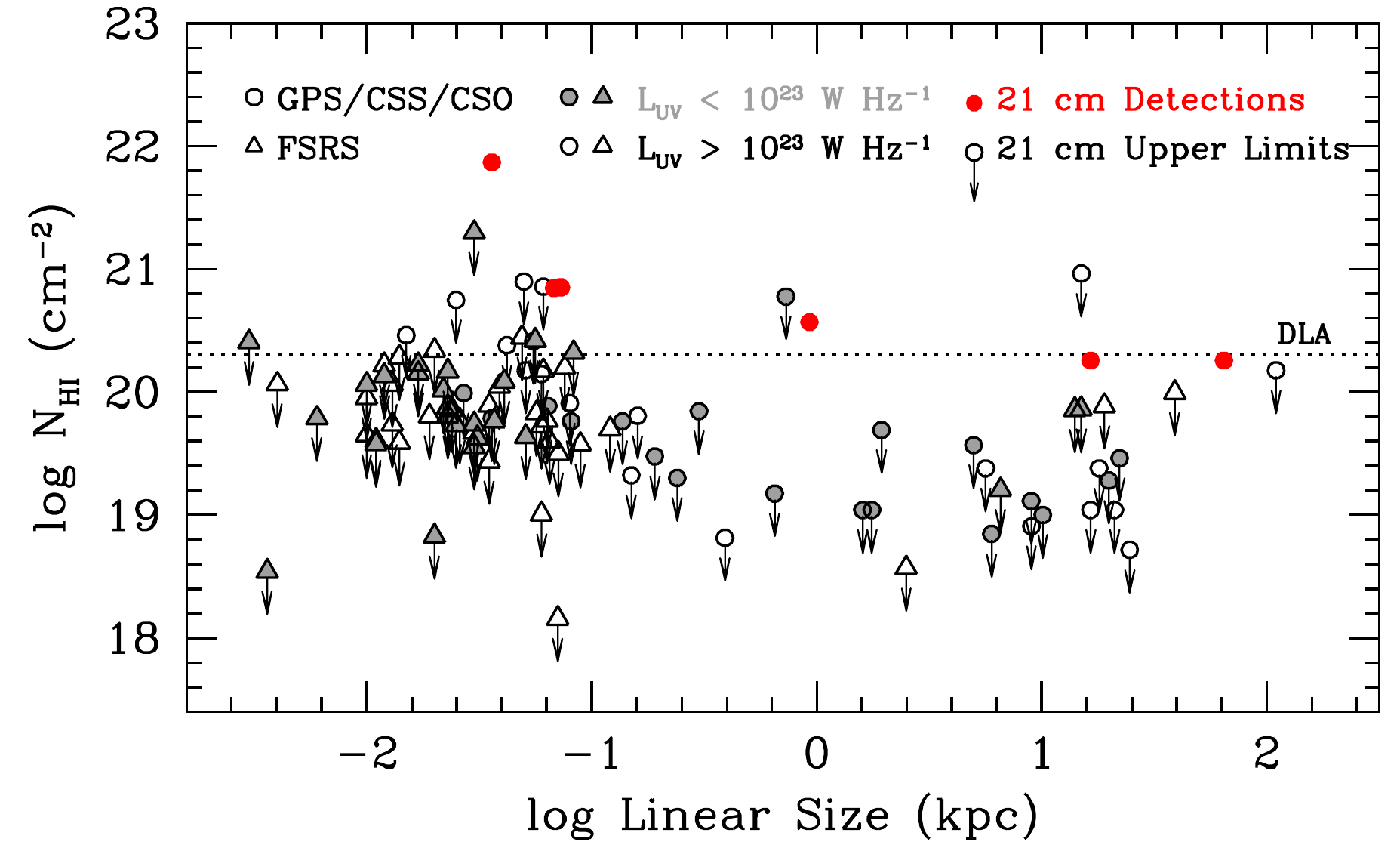}
\caption{HI column density (upper limit for sources not detected in HI)
  versus linear size of the radio source. The solid red
  symbols represent 21~cm detections, the filled gray data points represent non-detections
  with an intrinsic UV luminosity below the
  UV threshold, and black open symbols represent non-detections with a UV
  luminosity above the $10^{23}$~W~Hz$^{-1}$ threshold. The shape of
  the symbols represent the radio source identification: circles
  represent GPS, CSS, and CSO sources and triangles represent FSRSs.
  The horizontal dotted line indicates the threshold for damped Ly$\alpha$
  systems. 
\label{fig:NHIS}}
\end{figure}

It is becoming clear that absorption of cold, neutral gas within galaxies that host radio sources is a multifaceted problem, with many factors impacting the detection rate of HI 21~cm absorption, especially at high redshift. Different effects may influence each other and work together, such as the observed anti-correlation of the HI column densities with size of the radio source \citep{curran13, gereb15}. We do not recover the same correlation, although our data are not inconsistent with a decrease in the HI column density with increase in source size (Figure~\ref{fig:NHIS}). Our results are limited by an extremely small number of detections. Surveys such as this one, even with such a large non-detection rate, across a range of parameters (e.g., redshift, column density, UV/radio luminosity, radio source size) marks an attempt to better test and constrain the multitude of factors that impact the gas reservoirs and will allow for a more comprehensive understanding of the physical, kinematic, and chemical properties of the neutral gas in the AGN environment. 

Our low detection rate does not allow us to disentangle high (or even low) redshift biases against detection of 21~cm absorption. It is now clear that targeting UV-bright quasars, a condition imposed to reliably constrain the redshift of the source using the rest-frame UV continuum, are not identifying targets with large enough reservoirs of cold, neutral gas detectable with large radio surveys such as this study. Future surveys should be designed to target highly extincted, obscured sources that are not bright in the UV. Such surveys can leverage the infrared capabilities from WFIRST or even JWST to serendipitously detect potential HI targets that are UV/optical faint but bright in the mid-IR \citep[e.g.,][]{yan12, yan16}. The SKA and its pathfinders will undoubtedly inaugurate a new era in HI absorption surveys, owing to  large instantaneous bandwidth coverage and field of view. The improved sensitivity and spectral coverage achievable with the SKA will overcome observational biases introduced by the conventional requirement of an optical redshift, ushering in a future of true ``blind'' HI absorption surveys.

\section{Conclusions}\label{sec:conclusions}
We present the results of the largest survey to date to search for intrinsic HI 21~cm and OH 18~cm
absorption at redshifts between $0.02<z<3.8$ in 145 radio sources with the Green Bank Telescope. Below we summarize the major findings of this study.

\begin{enumerate}
\item Despite having the sensitivity to re-detect six previous detections of associated HI 21~cm absorption
in our sample, we make no new detections of 21~cm absorption in the
108 sources for which we have RFI-free spectra. The overall HI 21~cm absorption detection rate is 5.6\%, but no detections are made for $z > 0.7$. This is significantly lower than similar low redshift surveys. 

\item We place an
upper limit on the column density for all non-detections not
affected by RFI, reaching a mean optical depth of $\tau_{3\sigma}<0.023$. 79 of the radio sources have never been observed for HI 21~cm before. Despite low flux density values at the frequency of our
redshifted 21~cm spectral observations, coupled with the significant
presence of RFI in our low-frequency spectral observations, we 
reached adequate sensitivity to detect nearly any damped Ly$\alpha$ system. 

\item We have searched 125 of our sources for intrinsic molecular OH 18~cm
absorption. We do not detect any of the
four OH lines in the 102 RFI-free sources for which we have usable OH data,
obtaining exclusively upper limits for all measured OH column densities. We do not detect OH toward five of the known 21~cm absorption systems for which we have spectra coverage. Lack of OH absorption in these systems is not a result of insensitive measurements and we place upper limits on the OH column that are consistent with previously detected OH absorption systems. 

\item We find no HI 21~cm absorption above an ultraviolet luminosity of $L_{\rm UV} \gtrsim 10^{23}$~W~Hz$^{-1}$. This supports the hypothesis that intrinsic 21~cm absorption cannot be detected in galaxies when the UV luminosity exceeds this UV luminosity threshold \citep{curran08}. Lack of 21~cm absorption in UV bright sources is an effect due to either the ionization of the neutral gas, higher spin temperatures in the environment around these luminous AGN, or sources that simply do not have a sufficient column of cold, neutral HI along the line of sight. 

\item We do not find statistically significant evidence that the HI column density the of 21~cm systems in our sample depends on the radio jet size, UV luminosity, or radio (1.4~GHz) luminosity. We are limited, however, by small numbers. 

\item With the exception of a single source, all of our RFI-free sources beyond $z\sim1.2$ arise in UV-luminous sources above the UV luminosity threshold of $L_{\rm UV} = 10^{23}$~W~Hz$^{-1}$.
This suggests that a possible explanation that the dearth of intrinsic 21~cm absorption systems at high redshift is an selection bias favoring UV-luminous AGN. Requiring optical redshifts (rest-frame UV for $z\gtrsim 1.6$)
imposes an optical bias and low-dust column, selecting the brightest,
least obscured, and UV-bright radio sources at all redshifts. Such sources create an environment where neutral gas cannot survive and as such, are not detectable in 21~cm absorption.

\end{enumerate}

%Figure1

%figure5
\begin{figure}
\epsscale{1.15}
\plotone{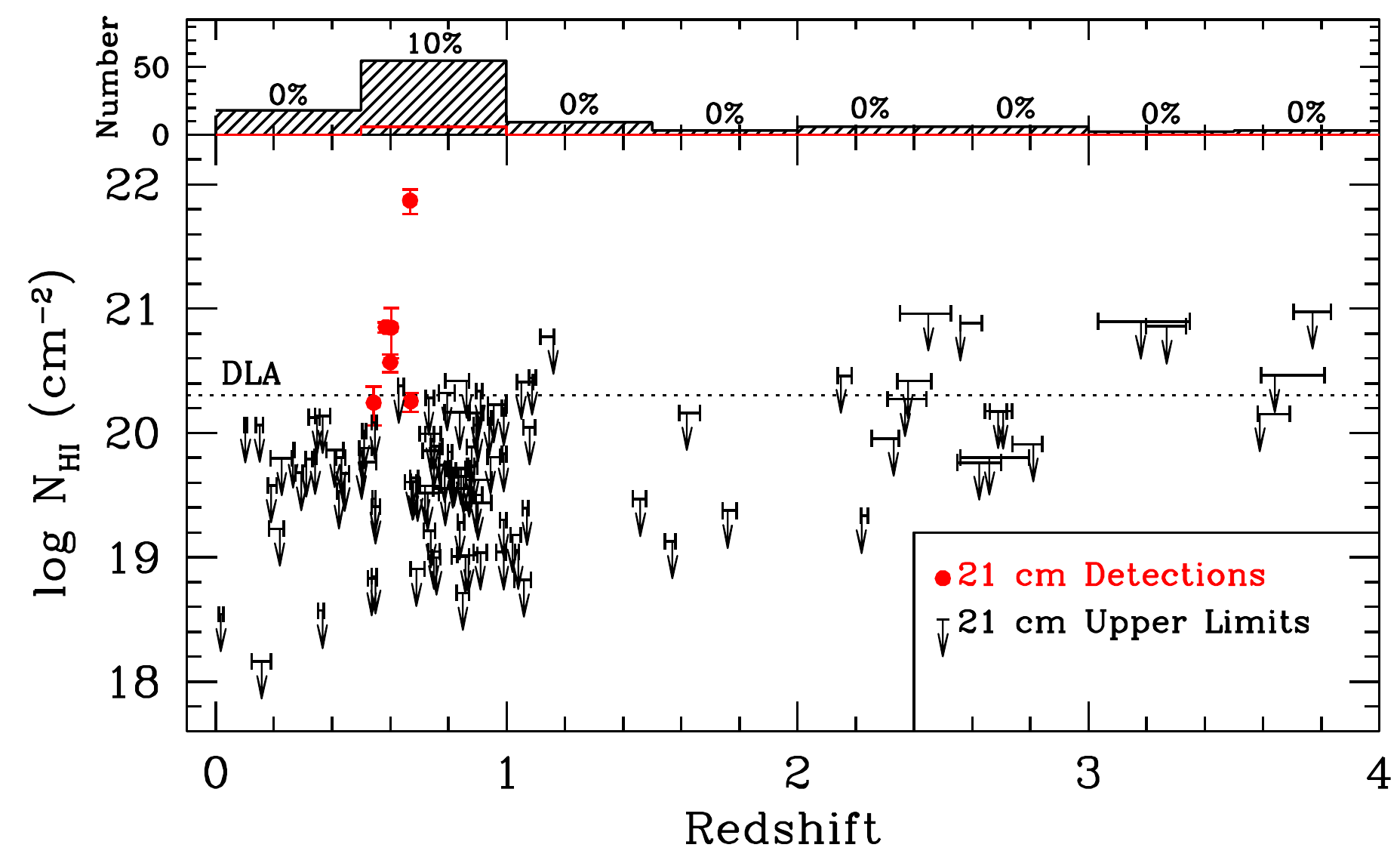}
\caption{HI column density versus redshift.  The lower panel shows either
  the detected HI column density at the redshift of the detected 21~cm absorption line or
  its 3$\sigma$ upper limit in the case of non-detections.
  The horizontal bar on the 21~cm non-detections indicates the redshift search
  region, and the downward-pointing arrow is centered on the systemic
  redshift of each object; the 21~cm line search regions are generally
  determined by RFI conditions. The solid red points represent sources
  detected in 21~cm absorption with the 1$\sigma$ error in the measured column
  density shown with the vertical error bars. The horizontal dotted line indicates the fiducial
  threshold for damped Ly$\alpha$ systems (DLAs; $N_{ \rm HI} \geq
  2\times10^{20}$~cm$^{-2}$). The upper panel
  indicates the number of objects surveyed in each $\Delta z=0.5$ redshift
  bin (black, hatched histogram), the number of intrinsic HI 21~cm
  absorption lines detected (red histogram), and the 21~cm absorption
  fraction (text percentages above each bin).   
\label{fig:NHIz}
}
\end{figure}

% figure6

\FloatBarrier

%figure7

%figure8

%figure9

%figure8

%figure11

%figure12

%Figure13

%Figure14

%Figure15

\acknowledgements
We are grateful for the valuable comments on this work by an anonymous referee that improved the scientific outcome and quality of the paper.
The authors thank T.\ Yan and K.\ Willett for data reduction scripts.  
The authors thank the staff members at the Green Bank Telescope for their assistance and support.
This research has made use of the NASA/IPAC Extragalactic Database (NED) which is operated by the Jet Propulsion Laboratory, California Institute of Technology, under contract with NASA.  
This publication makes use of data products from the Two Micron All Sky Survey, which is a joint project of the University of Massachusetts and the Infrared Processing and Analysis Center/California Institute of Technology, funded by the National Aeronautics and Space Administration and the National Science Foundation.
Parts of this research were supported by the Australian Research Council Centre of Excellence for All Sky Astrophysics in 3 Dimensions (ASTRO 3D), through project number CE170100013. The authors acknowledge the invaluable labor of the maintenance and clerical staff at their institutions, whose contributions make scientific discoveries a reality. 
KG acknowledges the staff at the University House for the supportive and collaborative writing environment.

{\it Facilities:}  \facility{GBT}

{\it Software:} GBTIDL \citep{marganian06}

%references

%table1
%780Mhz cntm comes from ctnminfo.dat
\LongTables
%%\startlongtable
\begin{deluxetable}{lllccccccc}
\tabletypesize{\scriptsize}
\tablecaption{Observations of Intrinsic HI and OH Lines Toward Compact Radio Sources\label{tab:table1}}
\tablecolumns{10}
\tablewidth{0pt}
\tablehead{
\colhead{Source} & 
\colhead{$\alpha$} & 
\colhead{$\delta$}  & 
\colhead{$z$} &
\colhead{$\nu_{\rm HI}$} &
\colhead{HI Time}  & 
\colhead{OH Time} & 
\colhead{Class.} &
\multicolumn{2}{c}{References}
\\
\colhead{} &
\colhead{(J2000)} &
\colhead{(J2000)} & 
\colhead{}  &
\colhead{(MHz)} &
\colhead{(sec)} &
\colhead{(sec)} &
\colhead{} &
\colhead{Class.} &
\colhead{Spec.} 
}
\startdata
%	Source		RA			DEC			z		nu		HI		OH	CLASS	REF	REF	
0018+729	&	00 21 27.4	&	+73 12 41.9	&	0.821(2)&	780.0(9)&	1493	&	896 	& GPS  	& 1  & \\
LBQS 0106+0119 	& 	01 08 38.8 	& 	+01 35 00.3 	& 	2.099(5)& 	472.1(8)&	564 	& 541	& FSRS	& 2	&\\
COINS J0111+3906&	01 11 37.3	&	+39 06 28.1	&	0.66847(3)	&	851.322(15)	&	1800	&\nodata& GPS,CSO  & 1,3,4  & 1\\
UM321 		& 	01 25 28.8 	& 	$-$00 05 55.9 	& 	1.07481(15)&	684.60(5)	&	298	&265	& FSRS	& 2	&\\
3C43		&	01 29 59.8	&	+23 38 20.3	&	1.459	&	577.6	&	1773	&	1725 	& CSS 	& 3,5,6,7 & \\
0133+476 	& 	01 36 58.6 	& 	+47 51 29.1 	& 	0.859 	&	764.1	&  	293  	& 290	& FSRS	& 2,8	& \\
3C48 		& 	01 37 41.3 	& 	+33 09 35.1 	& 	0.367 	&	1038	&	378    	& 366	& CSS	& 3,5,6	&\\       
3C49		&	01 41 09.2	&	+13 53 28.1	&	0.62	&	876.8	&	893	&	\nodata	& CSS 	& 3,5,6,7 & 2\\
\enddata
\tablecomments{Columns list the source name, Right Ascension and
  Declination, optical redshift, expected frequency of the corresponding HI
  21~cm line, total integration times for each reduced spectra, radio
  source classification (GPS, CSS, CSO, FSRS), references for the
  classification, and references for any previous HI 21~cm observations.
  Numbers in parentheses indicate uncertainties in the final digit(s) of
  listed quantities, when available. \\ 
Classification references: 
1 -- \cite{devries97};
2 -- \cite{healey07}; 
3 -- \cite{odea97}; 
4 -- \cite{peck00}; 
5 -- \cite{fanti90}; 
6 -- \cite{spencer89}; 
7 -- \cite{morganti97}; 
8 -- \cite{augusto06}; 
9 -- \cite{condon77}; 
10 -- \cite{xiang05}; 
11 -- \cite{jackson02};  
12 -- \cite{fomalont00}; 
13 -- \cite{gelderman94}. \\
Previous spectroscopy of the HI 21 cm line references:
1 -- \cite{carilli98};
2 -- \cite{ver03}; 
3 -- \cite{gupta06};
4 -- \cite{ishwara03};
5 -- \citet{curran13}.   
}
\end{deluxetable}

%Table2
\LongTables
%%\startlongtable
\begin{deluxetable}{lcccccccccc} 
\tabletypesize{\scriptsize}
\tablecaption{HI 21~cm Line Observations, Column Densities, and Source Sizes \label{tab:HI}}
\tablecolumns{11}
\tablewidth{0pt}
\tablehead{
\colhead{Source} & 
\colhead{$z$}  & 
\colhead{$\nu$} & 
\colhead{$\Delta v$}  &
\colhead{rms}  & 
\colhead{Ctnm}  &
\colhead{$N_{\rm HI}$} &
\colhead{Prev.\ $N_{\rm HI}$}  &
\multicolumn{2}{c}{Size} &
\colhead{Ref.}
\\		
\colhead{} & 
\colhead{} & 
\colhead{(MHz)} &
\colhead{(\kms)} &
\colhead{(mJy)} & 
\colhead{(Jy)} &
\multicolumn{2}{c}{($10^{20}$~cm$^{-2}$)} &
\colhead{(arcsec)} &
\colhead{(kpc)} &
\colhead{}
}
\startdata 
% Source	z		nu		del v		rms	 ctnm		N(HI)	pre N(HI)	ref	arc	kpc	ref	
0018+729	& 0.821(2)	& 780.0(9)	& $-$5077, 6080	& 3.37	& 1.0033(6)	& $<$0.51& \nodata & 0.018 & 0.14 & 1	\\
LBQS 0106+0119  & 2.009(5)	& 472.1(8)	& \nodata	& RFI	& \nodata	&\nodata & \nodata & 0.0106& 0.09 & 2	\\
COINS J0111+3906& 0.66847(3)	& 853.322(15)	& $-$1343, 3015	& 2.47	& 0.153(2)	& 74(13) & $81(17)^{1}$  & 0.005 & 0.04 & 3\\
UM321          	& 1.07481(15)   	& 684.60(5)    	& $-$3813, 26560& 8.3   & 1.306(18)	& $<$1.1 & \nodata & 0.0048& 0.039&4	\\
3C43		& 1.459		& 577.6		& $-$3047, 2470	& 12.8	& 5.97(4)	& $<$0.29& \nodata & 2.60 & 22.2 & 5\\
0133+476    	& 0.859		& 764.1 	& \nodata 	& RFI 	& \nodata	&\nodata & \nodata & 0.0029&0.022 & 2	\\
3C48       	& 0.367 	& 1038    	& $-$432, 568   & 4.9   & 20.04(4) 	& $<$0.037& \nodata & 0.50  & 2.5  & 5	\\
3C49		& 0.62		& 876.8		& \nodata	& RFI	& \nodata	&\nodata & $1.1^{2}$ & 0.99 & 6.7 & 3\\
\enddata
\tablecomments{Columns list the source name, optical redshift, centroid or expected 
  frequency for the HI 21~cm absorption line, velocity range searched (blue, red limits),
  spectral rms noise in the search region, the continuum flux 
  density at the expected redshifted line frequency, 
  the measured column density or $3\sigma$ upper limit, 
  previously measured column
  densities (or upper limits), largest projected
  angular size and linear size, respectively, and references for the continuum source characteristics.  Numbers in
  parentheses show uncertainties in the final digit(s). Detections and non-detections
  assume $f=1$, and $T_s=100$ K.  Non-detections also assume a line width of 30 km s$^{-1}$ (Sec.~\ref{sec:column}).\\ 
References for previous HI 21~cm observations (non-detections are converted to our standard linewidth and spin temperature to aid in comparison):  
1 -- \citet{carilli98};
2 -- \citet{ver03}; 
3 -- \citet{gupta06};
4 -- \citet{ishwara03};
5 -- \citet{curran13}; 
6 -- \citet{aditya18b}; 
7 -- \citet{aditya18a};
8 -- \citet{curran17};
9 -- \citet{curran19};
10 -- \citet{aditya16}.
References for sizes: 
1 -- \cite{stanghellini90}; 
2 -- \cite{gurvits99};
3 -- \cite{gupta06}; 
4 -- \cite{fomalont00}; 
5 -- \cite{odea97}; 
6 -- \cite{xu95}; 
7 -- \cite{fey00};  
8 -- \cite{menon83}; 
9 -- \cite{murphy93};
10 -- \cite{antonucci85};   
11 -- \cite{shen98};   
12 -- \cite{fanti01};  
13 -- \cite{fey97};  
14 -- \cite{xiang05};  
15 -- \cite{shen97};    
16 -- \cite{fanti90}; 
17 -- \citet{morganti93}.  
}
\end{deluxetable}

%\clearpage

%Table3
\clearpage
%\LongTables
%\begin{turnpage}
\begin{landscape}
%%\begin{longrotatetable}
%\begin{center}
\begin{deluxetable}{lcccccccccccccccc}
\tabletypesize{\scriptsize}
\tablecaption{OH 18~cm Observations and Column Density Limits
\label{tab:OH}}
\tablecolumns{13}
\tablewidth{0pt}
\tablehead{
\colhead{} &
\colhead{} &
\multicolumn{3}{c}{1612 MHz} & 
\colhead{} &
\multicolumn{3}{c}{1667 MHz} &
\colhead{} &
\multicolumn{3}{c}{1720 MHz} 
\\ \cline{3-5} \cline{7-9} \cline{11-13} \noalign{\smallskip}
\colhead{Source} & 
\colhead{$z$} & 
\colhead{$\nu$}  & 
\colhead{$\Delta v$} &
\colhead{rms} & 
\colhead{} &
\colhead{$\nu$}  & 
\colhead{$\Delta v$} &
\colhead{rms} & 
\colhead{} &
\colhead{$\nu$}  & 
\colhead{$\Delta v$} &
\colhead{rms} &
\colhead{Cntm} &
\colhead{$N_{\rm {OH}}/ T_x$}
\vspace{2pt}
\\
\colhead{} &
\colhead{} &
\colhead{(MHz)} & 
\colhead{(\kms)}  &
\colhead{(mJy)} &
\colhead{} &
\colhead{(MHz)} & 
\colhead{(\kms)}  &
\colhead{(mJy)} &
\colhead{} &
\colhead{(MHz)} & 
\colhead{(\kms)}  &   
\colhead{(mJy)} &   
\colhead{(Jy)} &
\colhead{($10^{13}$ cm$^{-2}$~K$^{-1}$)}
}
\startdata
0018+729	& 	0.821		& 	885.35		& 	\nodata	& 	RFI	& 		& 	915.63	& 	$-$8945, 9866	& 	2.41	& 		& 	944.83	& 	\nodata	& 	\nodata	& 	0.9244(5)	& 	$<$6.0\\
LBQS 0106+0119 		& 	2.009		& 	535.80		& 	$-$10741, 7163		& 	5.5		& 		& 	554.12		& 	$-$3179,	2231		& 	5.5		& 		& 	571.79		& 		$-$1681,	3562		& 		5.1		& 	2.7(2)		& 	$<$3.8	\\
UM321 			& 	1.07481	& 	777.05		& 	$-$4688, 3492		& 	9.1		& 		& 	803.62		& 	$-$4357,	4894		& 	6.1		& 		& 	829.25		& 		\nodata		& 		RFI		& 	1.355(18)		& 	$<$8.9	\\
3C43	& 	1.459	& 	655.64	& 	$-$153, 1868	& 	12.3	& 		& 	678.06	& 	$-$3188, 491	& 	19.1	& 		& 	699.69	& 	$-$1257, 2514	& 	18.2	& 	5.25(3)	& 	$<$8.5\\
0133+476 		& 	0.859		& 	867.26		& 	\nodata	& 	RFI		& 		& 	896.91		& 	$-$1834,	572		& 	3.9		& 		& 	925.51		& 	$-$1000,	5,900		& 	6.1		& 	1.983(11)		& 	$<$4.6	\\
3C48 			& 	0.367		& 	1179.4		& 	$-$3967, 3659		& 	8.1		& 		& 	1219.7		& 	$-$3755,	3618	& 		7.1		& 		& 	1258.6		& 	$-$5093,	623		& 	11.1		& 	17.93(4)		& 	$<$0.85	\\
PKS 0201+113	& 	3.6390		& 	347.54		& 	$-$3762, 7969	& 		9.78		& 		& 	359.42	& 	$-$4905, 3186		& 	10.1		& 		& 	370.88		& 		$-$10440, 5645		& 	11.9		& 	0.365(8)	& 	$<$41	\\
0212+735 		& 	2.367		& 	478.83		& 	\nodata	& 	RFI		& 		& 	495.21		& 	\nodata	& 	RFI		& 		& 		511.00		& 		\nodata		& 		RFI		& 	\nodata	& 	\nodata\\
\enddata
\tablecomments{Columns list the source name, the optical redshift, and 
  the expeted frequency, the veloctity range searched (blue, red limits), and the spectral rms noise 
  for each of the 1612, 1667, and 1720~MHz OH lines.  The final two columns list the continuum flux density 
  expected at the redshifted 1667 MHz OH line,  
  and the $3\sigma$ column density limit for that line, unless otherwise noted (Sec.\ \ref{sec:column}).  
  Numbers in parentheses show uncertainties in the final digit(s).  } 
\tablenotetext{a}{Since the 1667 (and 1665)~MHz line was not observable due
  to RFI, we use the 1612~MHz line to obtain an upper limit on the OH column density.
  The interpolated continuum corresponds to the redshifted frequency
  of this line. } 
\tablenotetext{b}{The interpolated continuum and OH column density limit
  are based on the 1720 MHz OH line. } 
\end{deluxetable}
%\end{turnpage}
%\end{center}
\clearpage 
\end{landscape}
%%\end{longrotatetable}

%Table4
\begin{deluxetable}{lccccccccc} 
\tabletypesize{\scriptsize}
\tablecaption{Intrinsic HI 21~cm Absorption Line Measurements and Derived Quantities\label{tab:detections}}
\tablecolumns{10}
\tablewidth{0pt}
\tablehead{
\colhead{Source} & 
\colhead{Comp.}  &
\colhead{$\nu$}  & 
\colhead{$z$} & 
\colhead{Ctnm}  &
\colhead{FWHM}  & 
\colhead{FWHM}  &
\colhead{Depth}  & 
\colhead{$\tau$}  &
\colhead{$N_{\rm HI}$} 
\\		
\colhead{} & 
\colhead{} &
\colhead{(MHz)} & 
\colhead{} &
\colhead{(Jy)}  &
\colhead{(kHz)} & 
\colhead{(\kms)} &
\colhead{(mJy)} & 
\colhead{($10^{-2}$)} &
\colhead{($10^{20}$~cm$^{-2}$)}
}
\startdata 
% source                nu            z            cntm     fwhm(khz) (km/s)  depth        	 tau           N_HI
COINS J0111+3906& 1 & 851.373(5) & 0.668371(10)& $0.153(2)$& 190(10)	& 67(4)  & 85(2)	& $81^{+2}_{-4}$ 	& $104^{+7}_{-8}$\\
		& 2 & 851.173(9) & 0.668763(18)& $0.153(2)$& 142(17)	& 50(6)  & 34(3)	& $25^{+2}_{-3}$ 	& $24^{+3}_{-4}$\\
                & Tot & 851.3278 & 0.6685      & 0.153(2)  & 180(20)   	& 63(7)  & 70(8)  	& 61(12) 	& 74(13)	\\  \\
4C+76.03        & 1 & 887.643(3) & 0.600199(5) & $6.618(7)$& 132(10)	& 45(3)  & 114(8)	& $1.74^{+0.12}_{-0.12}$ 	& $1.48^{+0.14}_{-0.15}$\\
		& 2 & 887.69(3)  & 0.60011(5)  & $6.618(7)$& 559(90)	& 189(29)& 36(6) 	& $0.53^{+0.09}_{-0.10}$ 	& $1.9^{+0.4}_{-0.5}$\\
                & Tot & 887.648  & 0.60019     & 6.618(7)  & 199(79)    & 67.2(27)& 165(6)   	& 2.9(1.1) 	& 3.7(1.1) \\
                \\
\enddata
\tablecomments{Sources for which 21~cm absorption is detected.  Columns
  list the source name, Gaussian component, central frequency, redshift,
  continuum flux density, FWHM width of the line (rest-frame), line depth, peak optical depth, and
  derived column density corresponding to each individual Gaussian component.  For rows labeled with Tot, these measurements correspond to the direct integration for each absorption profile.  
  Numbers in parentheses show statistical uncertainties in the final digit(s).} 
\end{deluxetable}

%\newpage

%Table5
%clearpage
%%\startlongtable
\begin{deluxetable}{lccccccccc} 
\tabletypesize{\scriptsize}
\tablecaption{UV, Optical, Near-IR, and Radio Properties\label{tab:magnitudes}}
\tablecolumns{10}
\tablewidth{0pt}
\tablehead{
\colhead{Source} & 
\colhead{$B$}  & 
\colhead{$V$}  & 
\colhead{$R$}  &
\colhead{$K$}  &
\colhead{$z$}  &
\colhead{$D_{\rm L}$} & 
\colhead{log $L_{\rm UV}$} &
\colhead{log $L_{\rm Radio}$} &
\colhead{References}
\\		
\colhead{} & 
\colhead{(mag)} & 
\colhead{(mag)} &
\colhead{(mag)} &
\colhead{(mag)} &
\colhead{} & 
\colhead{(Mpc)} &
\colhead{(W~Hz$^{-1}$)} &
\colhead{(W~Hz$^{-1}$)} &
\colhead{}
}
\startdata 
% source            	 B		V		R		K		z		D		logLuv	logLrad	Reference
0018+729	&	\nodata	&	\nodata	& 20.9(i-band)	&	\nodata &	0.821	&	5188.7	&	\nodata	& 27.25	&	1\\
COINS J0111+3906&	\nodata	&	\nodata	&	22.00	&	16.69	&	0.66847	&	4023.5	&	20.40	& 26.21	&	2,3\\
UM321          	& 	16.72	& 	16.44	& 	16.30	& 	14.41	& 	1.07481	& 	7303.7	& 	24.39	& 27.60	& 	4,5,6 \\
3C43		&	21.38	&	\nodata	&	19.70	&	20.0	&	1.459	&	10635	&	22.27	& 28.52	&	7,8\\
3C48       	& 	16.62	& 	16.2	& 	\nodata	& 	12.72	& 	0.367	& 	1950.7	& 	23.07	& 27.83	& 	7,6\\
0138$-$097   	& 	17.1	& 	\nodata	& 	17.0	& 	14.40	& 	0.733	& 	4508.2	& 	24.13	& 27.03	& 	10,6\\ 
PKS 0201+113	&	\nodata	&	\nodata	&	19.5	&	16.8	&	3.6390	&	32616	&	24.08	& 27.96	&	11\\
B2 0218+357  	& 	\nodata	& 	19.60	& 	\nodata &	14.84	& 	0.944	& 	6173.8	& 	22.38	& 27.56	& 	5,6\\
\enddata
\tablecomments{Columns list the source name, optical and near-IR magnitudes, luminosity 
  distance calculated according to \cite{wright06}, specific UV and 21~cm radio luminosity in the host galaxy
  reference frame, and references for the magnitudes. For sources with only
  one magnitude listed, we do not calculate a UV luminosity.  \\ 
Optical band magnitudes references:  
1 -- \citet{stanghellini93};
2 -- \citet{healey08}; 
3 -- \citet{stickel96}; 
4 -- \cite{wills78}; 
5 -- \cite{healey08}; 
6 -- 2MASS; \citet{skrutskie06}; 
7 -- \cite{snellen02}; 
8 -- SuperCOSMOS Sky Survey; \cite{hambly01}; 
9 -- \cite{sandage65}; 
10 -- \cite{atlee07}; 
11 -- \cite{kuhn04}; 
12 -- \cite{elvis94}; 
13 -- \cite{francis00}; 
14 -- \cite{kellermann98}; 
15 -- \cite{odell78a};
16 -- \cite{odell78}; 
17 -- \cite{glass81}; 
18 -- \cite{odea91}; 
19 -- \cite{sbarufatti05}; 
20 -- \cite{mcalary83}; 
21 -- \cite{godwin77}; 
22 -- \cite{gardner78}; 
23 -- \cite{wright83}; 
24 -- \citet{schmitt96}; 
25 -- \citet{drinkwater97}; 
26 -- \citet{cody03}; 
27 -- \cite{cur10}; 
28 -- \cite{sandage65a}; 
29 -- \citet{devries97b}; 
30 -- \cite{ellison05};  
31 -- \cite{lister05}; 
32 -- SDSS DR6; \citet{adelman08}; 
33 -- \cite{allen82}; 
34 -- \cite{ramirez04}; 
35 -- \cite{hyland82}; 
36 -- \cite{rao06}; 
37 -- \citet{chun06}; 
38 -- \cite{labiano07}; 
39 -- \citet{sbarufatti09}; 
40 -- \cite{chen05}; 
41 -- \citet{simpson00}; 
42 -- \cite{jorgenson06}; 
43 -- \citet{carballo99}; 
44 -- \citet{matthews63}; 
45 -- \cite{sandage65b}; 
46 -- \cite{fiorucci04}; 
47 -- \cite{tapia76}; 
48 -- \cite{kotilainen05}; 
49 -- \citet{ojha09}; 
50 -- \cite{devries00}; 
51 -- \cite{smith89}; 
52 -- \cite{drake04}; 
53 -- \citet{devries98}.     
}
\end{deluxetable}

\begin{figure*}
\epsscale{1.2}
\plotone{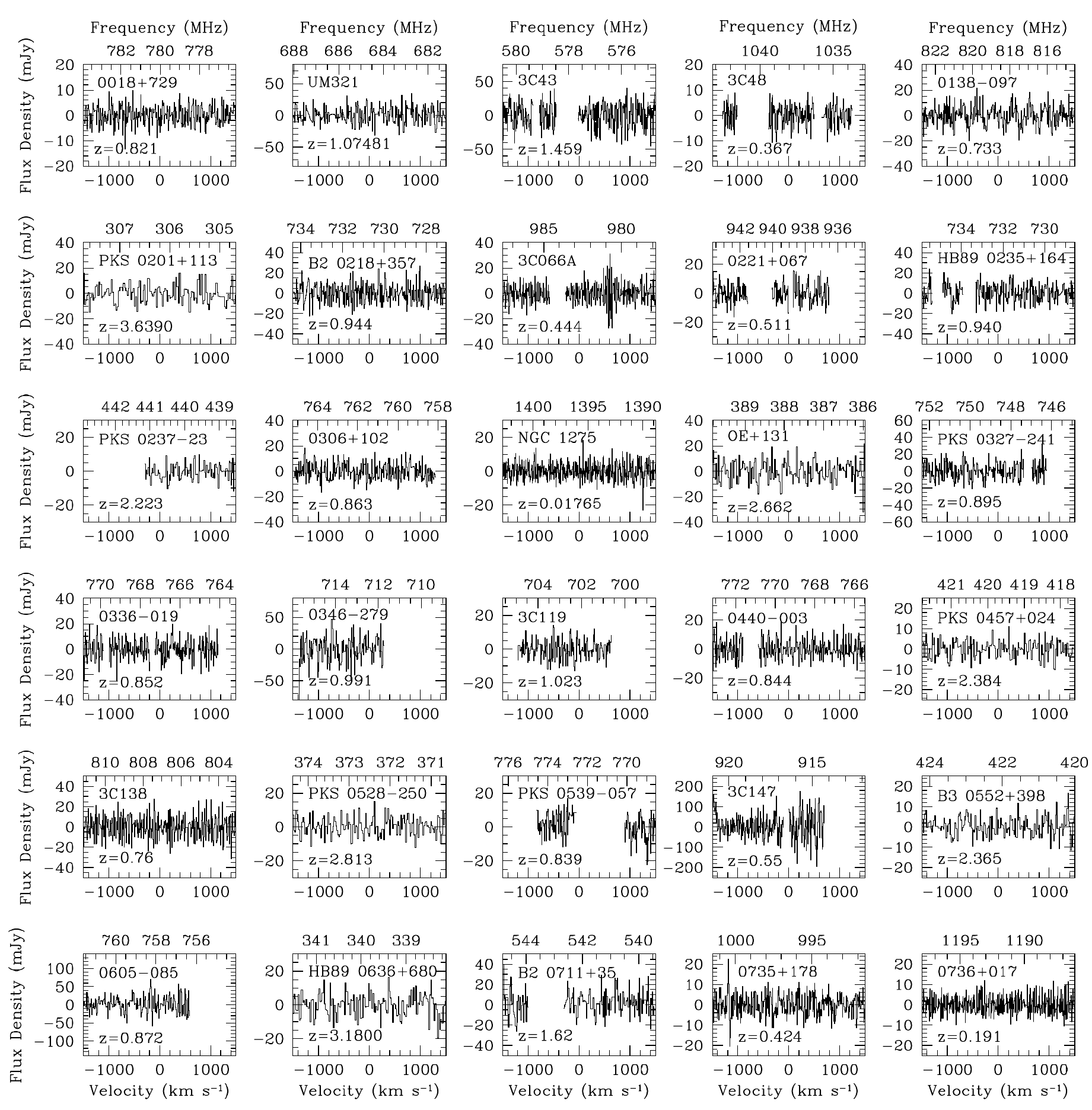}
\caption{ 
RFI-free sources not detected in HI 21~cm absorption.  The velocity scale
is in the rest frame of each object, defined by the heliocentric optical redshift of each radio
source (Table \ref{tab:table1}). Spectral regions lost to radio frequency
interference are not plotted.  Each spectrum spans 
$\pm1500$~\kms\ except for PKS~0742+10, where the x-axis has been shifted to show
the feature that arises at $z=2.64$. The upper spectrum in PKS~0742+10 shows the \citet{curran13}
data, supporting our non-detection interpretation of the possible absorption feature.  
\label{fig:nondetections}}
\end{figure*}

\begin{figure}
\epsscale{1.2}
\plotone{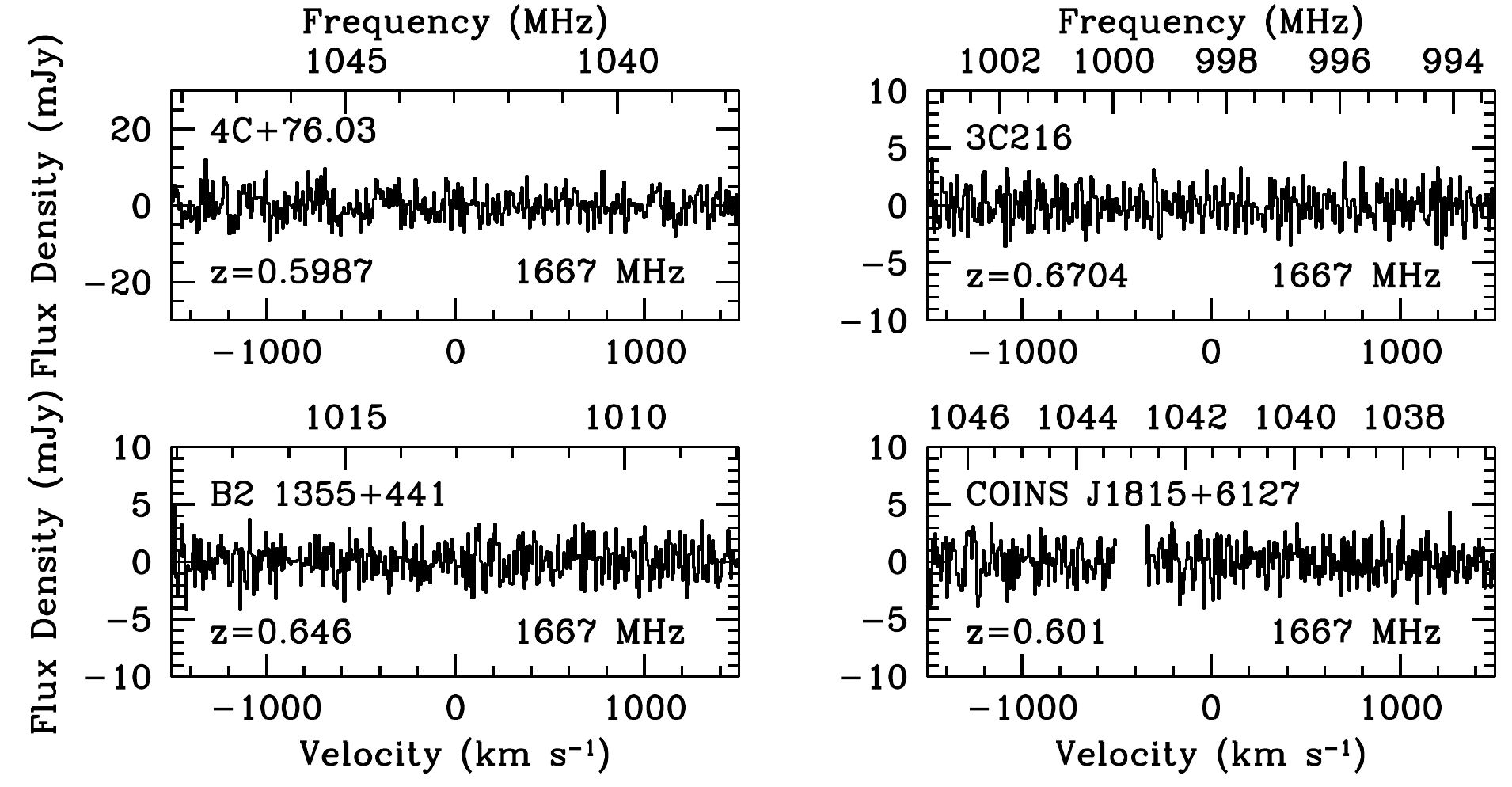}
\caption{ 
OH 1667 MHz spectra of the four RFI-free intrinsic HI 21~cm
absorption line systems searched for OH.  The velocity scale is in the
rest-frame of each source, defined by the redshift of the 21~cm absorber.  
Spectral regions lost to radio frequency interference are omitted.
\label{fig:OH_plot}}
\end{figure}

\begin{appendix}\label{sec:appendix}
We show the normalized spectra for the 102 RFI-free sources searched for intrinsic HI 21~cm (Figure~\ref{fig:nondetections}) and intrinsic OH 18~cm absorption in sources with known OH 21~cm absorption (Figure~\ref{fig:OH_plot}). The observable velocity span for each source is a few thousand \kms\ above/below with respect to the systemic velocity of the host galaxy and is typically determined by the RFI conditions rather than the spectral bandwidth. Table~\ref{tab:HI} lists the measured upper limits to the column densities for the 21~cm absorption line search. Table~\ref{tab:OH} lists the upper limits for the OH absorption line search. 

\end{appendix}

\begin{figure*}
\epsscale{1.2}
\figurenum{14}
%\plotone{plots1.eps}
%\plotone{plots1-eps-converted-to.pdf}
\plotone{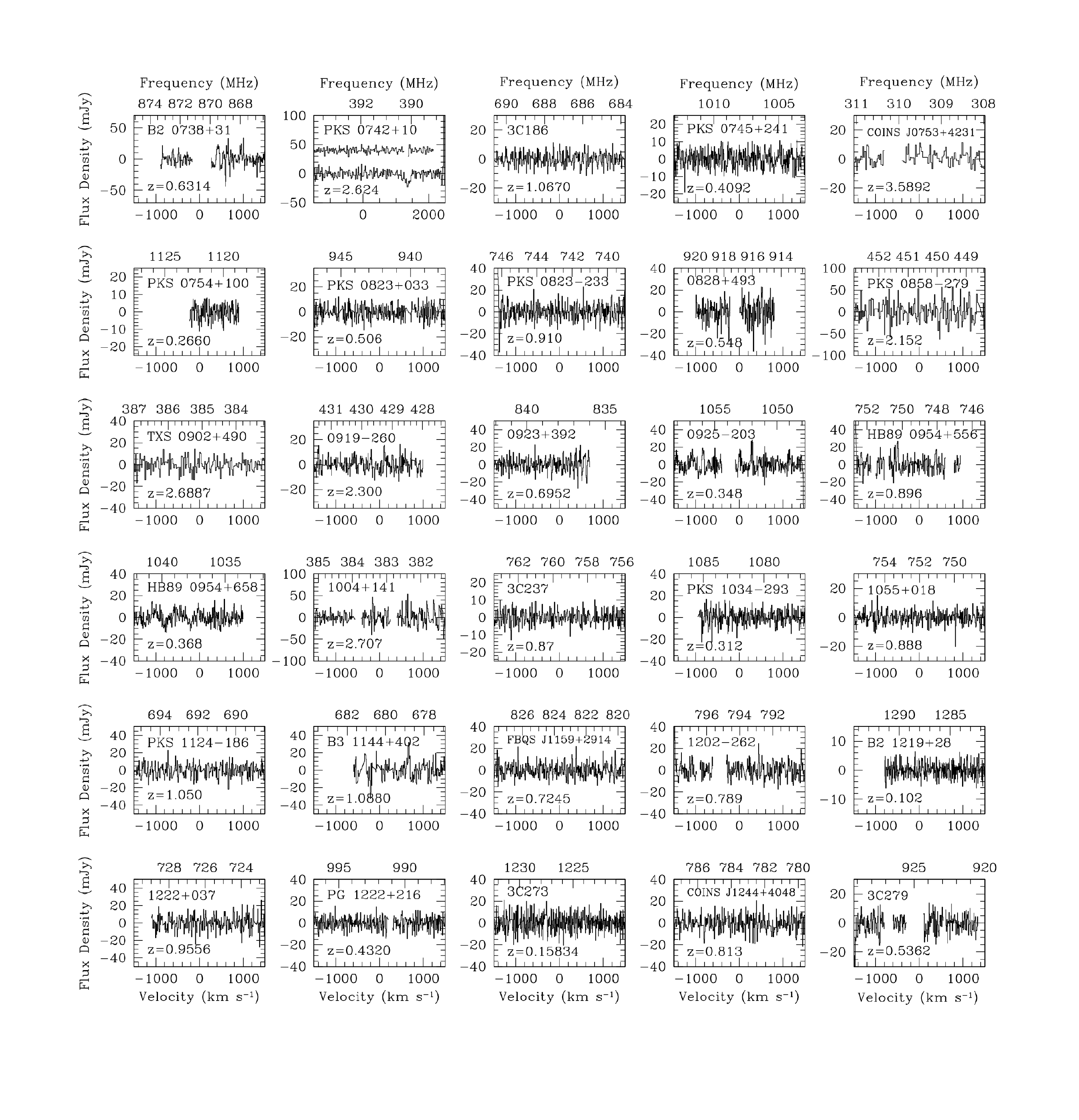}
\caption{con't.}
\end{figure*}

\begin{figure*}
\epsscale{1.2}
\figurenum{14}
\plotone{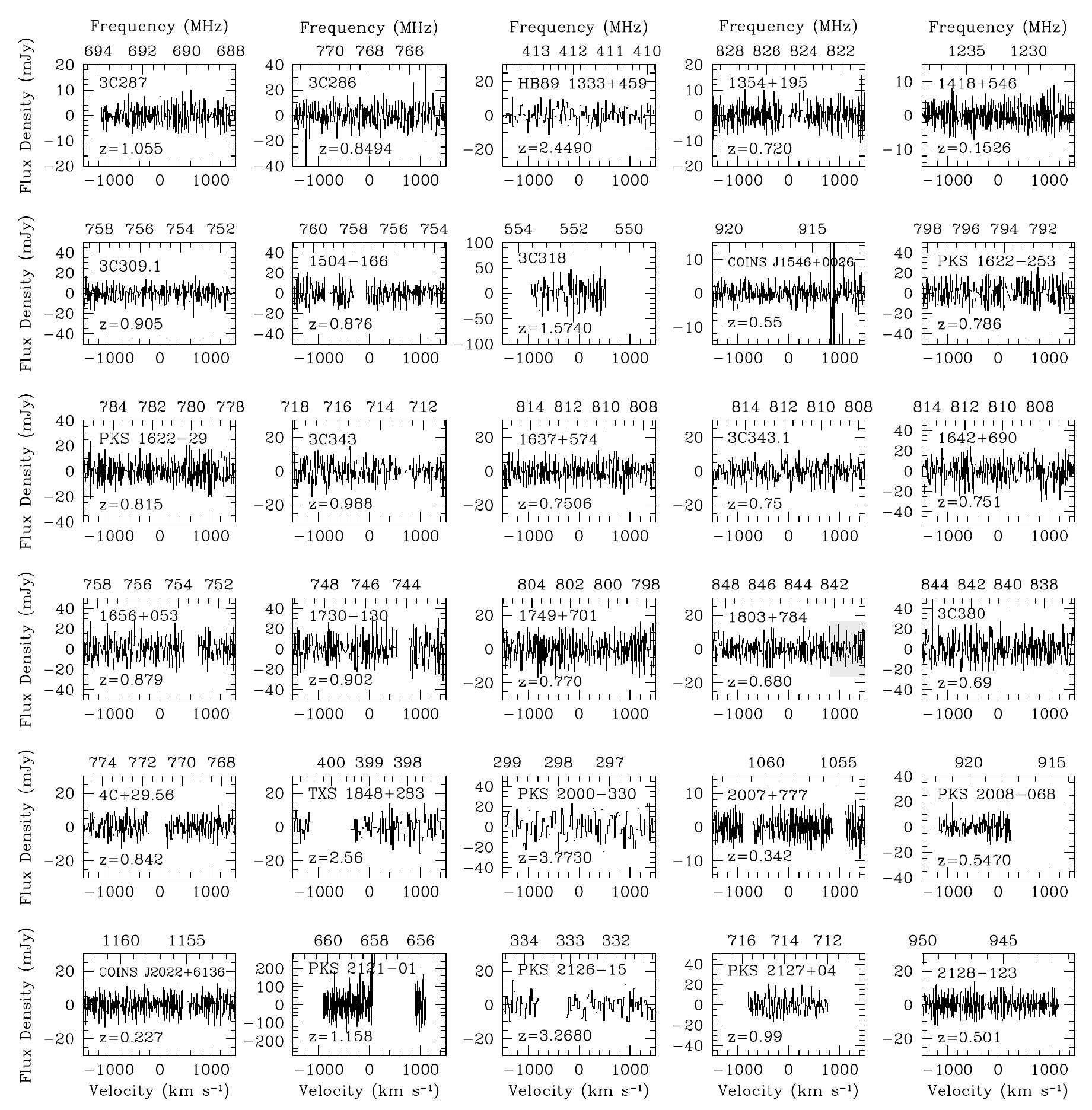}
\caption{con't.}
\end{figure*}

\begin{figure*}
\epsscale{1.2}
\figurenum{14}
\plotone{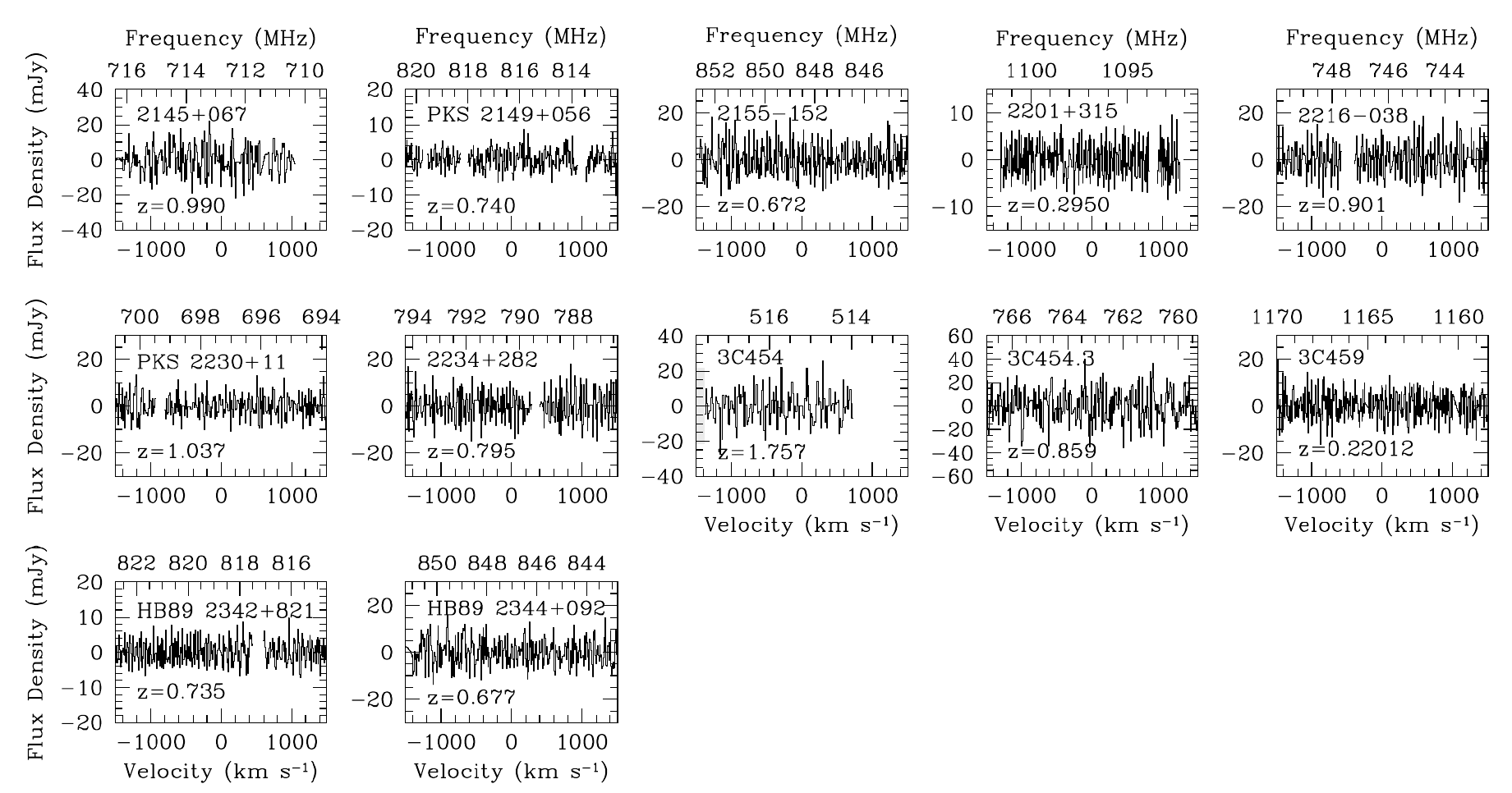}
\caption{con't.}
\end{figure*}

\newpage
\begin{thebibliography}{}
\bibitem[Adelman-McCarthy \etal(2008)]{adelman08} Adelman-McCarthy, J. K., Ag\"{u}eros, M. A., Allam, S. S. \etal\ 2008, \apjs, 175, 297
\bibitem[Aditya \etal(2016)]{aditya16} Aditya, J. N .H. S., Kanekar, N., \& Kurapati, S. 2016, \mnras, 455, 5000
\bibitem[Aditya \etal(2017)]{aditya17} Aditya, J. N .H. S., Kanekar, N., Prochaska, J. X., \etal\ 2017, \mnras, 465, 5011
\bibitem[Aditya \& Kanekar(2018a)]{aditya18a} Aditya, J. N. H. S. \& Kanekar, N. 2018a, \mnras, 473, 59
\bibitem[Aditya \& Kanekar(2018b)]{aditya18b} Aditya, J. N. H. S. \& Kanekar, N. 2018b, \mnras, 481, 1578
\bibitem[Allen, Ward \& Hyland(1982)]{allen82} Allen, D. A., Ward, M. J., \& Hyland, A. R. 1982, \mnras, 199, 969 
\bibitem[Antonucci \& Ulvestad(1985)]{antonucci85} Antonucci, R. R. J., \& Ulvestad, J. S. 1985, \apj, 294, 158  
\bibitem[Atlee \& Gould(2007)]{atlee07} Atlee, D. W. \& Gould, A. 2007, \apj, 664, 53 
\bibitem[Augusto \etal(2006)]{augusto06} Augusto, P., Gonzalez-Serrano, J. I., Perez-Fournon, I., \& Wilkinson, P. N. 2006, \mnras, 368, 1411  
\bibitem[Bajtlik \etal(1988)]{baj88} Bajtlik, S., Duncan, R. C., Ostriker, J. P., 1988, \apj, 327, 570  
\bibitem[Bessell(1979)]{bessell79} Bessell, M. S. 1979, \pasp, 91, 589
\bibitem[Campins \etal(1985)]{campins85} Campins, H., Rieke, G. H., \& Lebofsky, M. J. 1985, \aj, 90, 896
\bibitem[Carballo \etal(1999)]{carballo99} Carballo, R., Gonz\'{a}lez-Serrano, J. I, Benn, C. R., S\'{a}nchez, S. F., \& Vigotti, M. 1999, \mnras, 306, 137
\bibitem[Carilli \etal(1998)]{carilli98} Carilli, C. L., Menten, K. M., Reid, M. J., Rupen, M. P., \& Yun, M. S. 1998, \apj, 498, 175 
\bibitem[Chen \etal(2005)]{chen05} Chen, P. S., Fu, H. W., \& Gao, Y. F. 2005, NewA, 11, 27 
\bibitem[Chengalur \etal(1999)]{chengalur99} Chengalur, J. N., de Bruyn, A. G., \& Narasimha, D. 1999, ASPC, 156, 228 
\bibitem[Chengalur \& Kanekar(2003)]{chengalur03} Chengalur, J. N. \& Kanekar, N. 2003, Phys. Rev. Lett., 91, 241302 
\bibitem[Chun \etal(2006)]{chun06} Chun, M. R., Gharanfoli, S., Kulkarni, V. P., \& Takamiya, M. 2006, \aj, 131, 686
\bibitem[Cody \& Braun(2003)]{cody03} Cody, A. M. \& Braun, R. 2009, \aap, 400, 871 
\bibitem[Condon \etal(1977)]{condon77} Condon, J. J., Hicks, P. D, \& Jauncey, D. L. 1977, \aj, 82, 692 
\bibitem[Condon \etal(1998)]{condon98} Condon, J. J., Cotton, W. D., Greisen, E. W., \& Yin, Q. F. 1998, \aj, 115, 1693 
\bibitem[Cowie \& Songaila(1995)]{cowie95} Cowie, L. L. \&  Songaila, A. 1995, \apj, 453, 596
\bibitem[Curran \etal(2006)]{curran06} Curran, S. J., Whiting, M. T., Murphy, M. T. 2006, \mnras, 371, 431
\bibitem[Curran \etal(2008)]{curran08} Curran, S. J., Whiting, M. T., Wiklind, T., Webb, J. K, Murphy, M. T., \& Purcell, C. R. 2008, \mnras, 391, 765  
\bibitem[Curran \& Whiting(2010)]{cur10} Curran, S. J. \& Whiting, M. T. 2010, \apj, 712, 303 
\bibitem[Curran(2010)]{curran10} Curran, S. J. 2010, \mnras, 402, 2647
\bibitem[Curran \etal(2011)]{curran11} Curran, S. J., Whiting, M. T., Murphy, M. T., \etal\ 2011, \mnras, 413, 1165
\bibitem[Curran \etal(2013)]{curran13} Curran, S. J., Whiting, M. T., Sadler, E. M., \& Bignell, C. 2013, \mnras, 428, 2053
\bibitem[Curran \etal(2016)]{curran16} Curran, S. J., Duchesne, S. W., Divoli, A., \& Allison, J. R. 2016, \mnras, 462, 4197
\bibitem[Curran \etal(2017)]{curran17} Curran, S. J., Hunstead, R. W., Johnston, H. M., \etal\ 2017, \mnras, 470, 4600
\bibitem[Curran(2018)]{curran18} Curran, S. J. 2018, PASA, 35, 36
\bibitem[Curran \etal(2019)]{curran19} Curran, S. J., Hunstead, R. W., Johnston, H. M., \etal\ 2019, \mnras, 484, 1182
\bibitem[Darling(2003)]{darling03} Darling, J. 2003, Phys. Rev. Lett., 91, 011301
\bibitem[Darling(2004)]{darling04} Darling, J. 2004, \apj, 612, 58    
\bibitem[de Vries, Barthel, \& O'Dea(1997)]{devries97} de Vries, W. H., Barthel, P. D., \& O'Dea, C. P. 1997, \aap, 321, 105 
\bibitem[de Vries \etal(1997b)]{devries97b} de Vries, W. H., O'Dea, C. P., Baum, S. A., \etal\ 1997b, \apjs, 110, 191
\bibitem[de Vries \etal(1998)]{devries98} de Vries, W. H., O'Dea, C. P., Perlman, E., Baum, S. A., Lehnert, M. D., Stocke, J., Rector, T., \& Elston, R. 1998, \apj, 503, 138
\bibitem[de Vries \etal(2000)]{devries00} de Vries, W. H., O'Dea, C. P., Barthel, P. D., \etal\ 2000, \apj, 120, 2300 
\bibitem[Douglas \etal(1996)]{douglas96} Douglas, J. N., Bash, F. N., \& Bozyan, F. A. 1996, \aj, 111, 1954 
\bibitem[Drake \etal(2004)]{drake04} Drake, C. L., McGregor, P. J., \& Dopita, M. A. 2004, \aj, 128, 955 
\bibitem[Drinkwater \etal(1997)]{drinkwater97} Drinkwater, M. J., Webster, R. L., Francis, P. J., \etal\ 1997, \mnras, 284, 85 
\bibitem[Elitzur(1992)]{elitzur92} Elitzur, M. 1992, Astronomical Masers (Dordrecht: Kluwer); http://adsabs.harvard.edu/abs/1992ASSL..170.....E
\bibitem[Ellison, Hall, \& Lira(2005)]{ellison05} Ellison, S. L., Hall, P. B., \& Lira, P. 2005, \aj, 130, 1345
\bibitem[Elvis \etal(1994)]{elvis94} Elvis, M., Wilkes, B. J., McDowell, J. C., \etal\ 1994, \apjs, 95, 1
\bibitem[Fanti \etal(1989)]{fanti89} Fanti, C., Fanti, R., Parma, P., \etal\ 1989, \aap, 217, 44
\bibitem[Fanti \etal(1990)]{fanti90} Fanti, R., Fanti, C., Schilizzi, R. T., \etal\ 1990, \aap, 231, 333 
\bibitem[Fanti \etal(2001)]{fanti01} Fanti, C., Pozzi, F., Dallacasa, D., \etal\ 2001, \aap, 369, 380  
\bibitem[Fey \& Charlot(1997)]{fey97} Fey, A. L. \& Charlot, P. 1997, \apjs, 111, 95 
\bibitem[Fey \& Charlot(2000)]{fey00} Fey, A. L. \& Charlot, P. 2000, \apjs, 128, 17 
\bibitem[Ficarra, Grueff, \& Tomassetti(1985)]{ficarra85} Ficarra, A., Grueff, G., \& Tomassetti, G. 1985, \aaps, 59, 255
\bibitem[Field(1959)]{field59} Field, G. B. 1959, \apj, 129, 536
\bibitem[Fiorucci, Ciprini, \& Tosti(2004)]{fiorucci04} Fiorucci, M., Ciprini, S., \& Tosti, G. 2004, \aap, 419, 25
\bibitem[Fomalont \etal(2000)]{fomalont00} Fomalont, E. B., Frey, S., Paragi, Z., \etal\ H. 2000, \apjs, 131, 95 
\bibitem[Francis \etal(2000)]{francis00} Francis, P. J., Whiting, M. T., \& Webster, R. L. 2000, \pasp, 17, 56 
\bibitem[Fugmann(1988)]{fugmann88} Fugmann, W. 1988, \aap, 205, 86
\bibitem[Gardner \& Whiteoak(1978)]{gardner78} Gardner, F. F. \& Whiteoak, J. B. 1978, \mnras, 183, 711 
\bibitem[Gelderman \& Whittle(1994)]{gelderman94} Gelderman, R. \& Whittle, M., 1994, \apjs, 91, 491 
\bibitem[Ger\'{e}b \etal(2014)]{gereb14} Ger\'{e}b, K., Morganti, R., \& Oosterloo, T. A. 2014, \aap, 569, A35
\bibitem[Ger\'{e}b \etal(2015)]{gereb15} Ger\'{e}b, K., Maccagni, F. M., Morganti, R., \& Oosterloo, T. A. 2015, \aap, 575, 44
\bibitem[Glass(1981)]{glass81} Glass, I. S. 1981, \mnras, 194, 795
\bibitem[Godwin \etal(1977)]{godwin77} Godwin, J. G. Bucknell, M. J., Dixon, K. L., Green, M. R., Peach, J. V., \& Wallis, R. E. 1977, Obs, 97, 238 
\bibitem[Gupta \etal(2006)]{gupta06} Gupta, N., Salter, C. J., Saikia, D. J., Ghosh, T., \& Jeyakumar, S. 2006, \mnras, 373, 972 
\bibitem[Gurvits \etal(1999)]{gurvits99} Gurvits, L. I.,  Kellermann, K. I.,  \& Frey, S. 1999, \aap, 342, 378 
\bibitem[Hambly \etal(2001)]{hambly01} Hambly, N., MacGillivray, H. T., Read, M. A., \etal\ 2001, \mnras, 326, 1279 
\bibitem[Healey \etal(2007)]{healey07} Healey, S. E., Romani, R. W., Taylor, G. B., \etal\ 2007, \apjs, 171, 61 
\bibitem[Healey \etal(2008)]{healey08} Healey, S. E., Romani, R. W., Cotter, G., \etal\ 2008, \apjs, 175, 97 
\bibitem[Hyland \& Allen(1982)]{hyland82} Hyland, A. R. \& Allen, D. A. 1982, \mnras, 199, 943 
\bibitem[Ishwara-Chandra \etal(2003)]{ishwara03} Ishwara-Chandra, C. H., Dwarakanath, K. S. \& Anantharamaiah, K. R. 2003, JA\&A, 24, 37 
\bibitem[Jackson \etal(2002)]{jackson02} Jackson, C. A., Wall, J. V., Shaver, P. A., Kellermann, K. I., Hook, I. M., \& Hawkins, M. R. S. 2002, \aap, 386, 97 
\bibitem[Jones \etal(1974)]{jones74} Jones, T. W., O'Dell, S. L., \& Stein, W. A. 1974, \apj, 192, 261 
\bibitem[Jorgenson \etal(2006)]{jorgenson06} Jorgenson, R. A., Wolfe, A. M., Prochaska, J. X., \etal\ 2006, \apj, 646, 730  
\bibitem[Kanekar \etal(2003)]{kanekar03} Kanekar, N., Chengalur, J. N., de Bruyn, A. G., \& Narasimha, D. 2003, \mnras, 345, 7 
\bibitem[Kanekar \& Chengalur(2003)]{kanekarchengular03} Kanekar, N. \& Chengalur, J. N. 2003, \aap, 399, 857
\bibitem[Kanekar \& Chengalur(2004)]{kanekar04} Kanekar, N. \& Chengalur, J. N. 2004, \mnras, 350, 17
\bibitem[Kanekar \etal(2005)]{kanekar05} Kanekar, N., Carilli, C. L., Langston, G. I., \etal\ 2005, Phys. Rev. Lett., 95, 261301 
\bibitem[Kanekar \etal(2006)]{kanekar06} Kanekar, N., Subrahmanyan, R., Ellison, S. L., \etal\ 2006, \mnras, 370, L46
\bibitem[Kanekar \etal(2012)]{kanekar12} Kanekar, N., Langston, G. I., Stocke, J. T., Carilli, C. L. \& Menten, K. M. 2012, \apjl, 746, L16
\bibitem[Kanekar \etal(2014)]{kanekar14a} Kanekar, N., Prochaska, J. X., Smette, A., \etal\ 2014, \mnras, 438, 2131
\bibitem[Kanekar \etal(2018)]{kanekar18} Kanekar, N., Ghosh, T., \& Chengalur, J. N. 2018, Phys. Rev. Lett., 120, 061302
\bibitem[Kellermann \etal(1998)]{kellermann98} Kellermann, K. I., Vermeulen, R. C., Zensus, J. A., \& Cohen, M. H. 1998, \aj, 115, 1295 
\bibitem[Komatsu \etal(2011)]{komatsu11} Komatsu, E., Smith, K. M., Dunkley, J., \etal\ 2011, \apjs, 192, 18 
\bibitem[Kotilainen \etal(2005)]{kotilainen05} Kotilainen, J. K., Hyv\"{o}nen, T., \& Falomo, R. 2005, \aap, 440, 831    
\bibitem[Kuhn(2004)]{kuhn04} Kuhn, O. P. 2004, \mnras, 348, 647
\bibitem[Labiano \etal(2007)]{labiano07} Labiano, A., Barthel, P. D., O'Dea, C. P., de Vries, W. H., P\'{e}rez, I., \& Baum, S. A. 2007, \aap, 463, 97   
\bibitem[Laing \& Peacock(1980)]{laing80} Laing, R. A. \& Peacock, J. A. 1980, \mnras, 190, 903 
\bibitem[Large \etal(1981)]{large81} Large, M. I., Mills, B. Y., Little, A. G., Crawford, D. F., \& Sutton, J. M. 1981, \mnras, 194, 693   
\bibitem[Lister \& Homan(2005)]{lister05} Lister, M. L. \& Homan, D. C. 2005, \aj, 130, 1389 
\bibitem[Maccagni \etal(2017)]{maccagni17} Maccagni, F. M., Morganti, R., Oosterloo, T. A., Ger\'{e}b, K., \& Maddox, N. 2017, \aap, 604, A43
\bibitem[Marganian \etal(2006)]{marganian06} Marganian, P., Garwood, R. W., Braatz, J. A., Radziwill, N. M., \& Maddalena, R. J. 2006, ASPC, 351, 512
\bibitem[Matthews \& Sandage(1963)]{matthews63} Matthews, T. A. \& Sandage, A. R. 1963, \apj, 138, 30
\bibitem[McAlary \etal(1983)]{mcalary83} McAlary, C. W., McLaren, R. A., McGonegal, R. J. \& Maza, J. 1983, \apjs, 52, 341 
\bibitem[Menon(1983)]{menon83} Menon, T. K. 1983, \aj, 88, 598
\bibitem[Molina \etal(2012)]{molina12} Molina, M., Landi, R., Bassini, \etal\ 2012, \aap, 548, 32
\bibitem[Moore \etal(1999)]{moore99} Moore, C. B., Carilli, C. L., \& Menten, K. M. 1999, \apj, 510, L87
\bibitem[Morganti, Killeen, \& Tadhunter(1993)]{morganti93} Morganti, R., Killeen, N. E. B., \& Tadhunter, C. N., 1993, \mnras, 263, 1023
\bibitem[Morganti \etal(1997)]{morganti97} Morganti, R., Oosterloo, T. A., Reynolds, J. E., \& Tadhunter, C. N.  1997, \mnras, 284, 541
\bibitem[Morganti \etal(2001)]{morganti01} Morganti, R., Oosterloo, T. A., Tadhunter, C. N., van Moorsel, G., Killeen, N., \& Wills, K. A. 2001, \mnras, 323, 331 
\bibitem[Moss \etal(2017)]{moss17} Moss, V. A., Allison, J. R., Sadler, E. M., \etal\ 2017, \mnras, 471, 2952
\bibitem[Murphy \etal(1993)]{murphy93} Murphy, D. W., Browne, I. W. A., \& Perley, R. A. 1993, \mnras, 264, 298  
\bibitem[Murphy \etal(2001)]{murphy01} Murphy, M. T., Webb, J. K., Flambaum, V. V., \etal 2001, \mnras, 327, 1208
\bibitem[O'Dea \etal(1991)]{odea91} O'Dea, C. P., Baum, S. A., \& Stanghellini, C. 1991, \apj, 380, 660 
\bibitem[O'Dea \& Baum(1997)]{odea97} O'Dea, C. P. \& Baum, S. A. 1997, \apj, 113, 148
\bibitem[O'Dell \etal(1978)]{odell78} O'Dell, S. L., Puschell, J. J., Stein, W. A., \& Warner, J. W. 1978, \apjs, 38, 267
\bibitem[Odell \etal(1978a)]{odell78a} Odell, S. L., Puschell, J. J., Stein, W. A., \etal\ 1978a, \apj, 224, 22 
\bibitem[Ojha \etal(2009)]{ojha09} Ojha, R., Zacharias, N., Hennessy, G. S., Gaume, R. A., \& Johnston, K. J. 2009, \aj, 138, 845
\bibitem[Orienti \etal(2007)]{orienti07} Orienti, M., Dallacasa, D., \& Stanghellini, C. 2007, \aap, 475, 813
\bibitem[Owsianik \& Conway(1998)]{owsianik98} Owsianki, I. \& Conway, J. E. 1998, \aap, 337, 69 
\bibitem[Peacock \& Wall(1982)]{peacock82} Peacock, J. A. \& Wall, J. V. 1982, \mnras, 198, 843 
\bibitem[Peck \& Taylor(2000)]{peck00} Peck, A. B. \& Taylor, G. B. 2000, \apj, 534, 90 
\bibitem[Petrov \etal(2008)]{petrov08} Petrov, L., Kovalev, Y. Y., Fomalont, E. B., \& Gordon, D. 2008, \aj, 136, 580 
\bibitem[Pihlstr\"{o}m \etal(2003)]{pih03} Pihlstr\"{o}m, Y. M., Conway, J. E., \& Vermeulen, R. C. 2003, \aap, 404, 871
\bibitem[Rahmani \etal(2012)]{rahmani12} Rahmani, H., Srianand, R., Gupta, N., \etal\ 2012, \mnras, 425, 556
\bibitem[Ram\'{i}rez \etal(2004)]{ramirez04} Ram\'{i}rez, A., de Diego, J. A, Dultzin-Hacyan, D., \& Gonz\'{a}lez-P\'{e}rez, J. N. 2004, \aap, 421, 83 
\bibitem[Rao \etal(2006)]{rao06} Rao, S. A., Turnshek, D. A, \& Nestor, D. B. 2006, \apj, 636, 610 
\bibitem[Readhead \etal(1996)]{readhead96} Readhead, A. C. S, Taylor, G. B., Xu, W., Pearson, T. J., Wilkinson, P. N., \& Polatidis, A. G. 1996, \apj, 460, 612 
\bibitem[Rengelink \etal(1997)]{renge97} Rengelink, R .B, Tang, Y., de Bruyn, A. G., \etal\ 1997, \aap, 124, 259 
\bibitem[Rhee \etal(2018)]{rhee18} Rhee, J., Lah, P., Briggs, F. H., \etal\ 2018, \mnras, 473, 1879
\bibitem[Roberts(1970)]{roberts70} Roberts, M. S, 1970, \apj, 161, L9 
\bibitem[Sandage(1965)]{sandage65} Sandage, A. 1965, \apj, 141, 1560
\bibitem[Sandage \etal(1965a)]{sandage65a} Sandage, A., Philippe, V., \& Wyndham, J. D. 1965a, \apj, 142, 1307 
\bibitem[Sandage \& Wyndham(1965b)]{sandage65b} Sandage, A. \& Wyndham, J. D. 1965b, \apj, 141, 328 
\bibitem[Sbarufatti \etal(2005)]{sbarufatti05} Sbarufatti, B., Treves, A., Falomo, R., Heidt, J., Kotilainen, J., \& Scarpa, R. 2005, \aj, 129, 559
\bibitem[Sbarufatti \etal(2009)]{sbarufatti09} Sbarufatti, B.. Ciprini, S., Kotilainen, J., \etal\ 2009, \aj, 137, 337
\bibitem[Schmitt \& Kinney(1996)]{schmitt96} Schmitt, H. R. \& Kinney, A. L. 1996, \apj, 463, 498
\bibitem[Schneider \etal(1983)]{schneider83}  Schneider, D. P., Gunn, J. E., \& Hoessel, J. G. 1983, \apj, 264, 337 
\bibitem[Shen \etal(1997)]{shen97} Shen, Z. -Q., Wan, T. -S., Moran, J. M., Jauncey, D. L., Reynolds, J. E., \& Tzioumis, A. K. 1997, \aj, 114, 1999   
\bibitem[Shen \etal(1998)]{shen98} Shen, Z. -Q., Wan, T. -S., Moran, J. M., \etal\ 1998, \aj, 115, 1357 
\bibitem[Simpson \& Rawlings(2000)]{simpson00} Simpson, C. \& Rawlings, S. 2000, \mnras, 317, 1023
\bibitem[Skrutskie \etal(2006)]{skrutskie06} Skrutskie, M. F., Cutri, R. M., Stiening, R., \etal\ 2006, \aj, 131, 1163 
\bibitem[Smith \& Heckman(1989)]{smith89} Smith, E. P. \& Heckman, T. M. 1989, \apjs, 69, 365 
\bibitem[Smith \etal(2002)]{smith02} Smith, J. A., Tucker, D. L., Kent, S., \etal\ 2002, \aj, 123, 2121 
\bibitem[Snellen \etal(2002)]{snellen02} Snellen, A. G., Lehnert, M. D., Bremer, M .N., \& Schilizzi, R. T. 2002, \mnras, 337, 981 
\bibitem[Spencer \etal(1989)]{spencer89} Spencer, R. E., McDowell, J. C., Charlesworth, M., Fanti, C., Parma, P., \& Peacock, J. A. 1989, \mnras, 240, 657 
\bibitem[Srianand \etal(2010)]{srianand10} Srianand, R., Gupta, N., Petitjean, P., Noterdaeme, P., \& Ledoux, C. 2010, \mnras, 405, 1888
\bibitem[Stanghellini \etal(1990)]{stanghellini90} Stanghellini, C., Baum, S. A., O'Dea, C. P., \& Morris, G. B. 1990, \aap, 233, 379 
\bibitem[Stanghellini \etal(1993)]{stanghellini93} Stanghellini, C., O'Dea, C. P., Baum, S. A.,  \& Laurikainen, E. 1993, \apjs, 88, 1 
\bibitem[Stanghellini \etal(1998)]{stanghellini98} Stanghellini, C., O'Dea, C. P., Dallacasa, D., Baum, S. A., Fanti, R., \& Fanti, C. 1998, \aaps, 131, 303
\bibitem[Stanghellini \etal(2005)]{stanghellini05} Stanghellini, C., O'Dea, C. P., Dallacasa, D., \etal\ 2005, \aap, 443, 891 
\bibitem[Stickel \& Kuehr(1996)]{stickel96} Stickel, M. \& Kuehr, H. 1996, \aaps, 115, 11
\bibitem[Tapia \etal(1976)]{tapia76} Tapia, S., Craine, E. R., \& Johnson, K. 1976, \apj, 203, 291
\bibitem[Tzanavaris \etal(2005)]{tzanavaris05} Tzanavaris, P, Webb, J. T., Murphy, M. T., Flambaum, V. V., \& Curran, S. J. 2005, Phys. Rev. Lett, 95, 041301
\bibitem[Tzanavaris \etal(2007)]{tzanavaris07} Tzanavaris, P, Murphy, M. T., Webb, J. T., Flambaum, V. V., \& Curran, S. J. 2007, \mnras, 374, 634
\bibitem[Uson \etal(1991)]{uson91} Uson, J. M., Bagri, D. S., \& Cornwell, T. J. 1991, Phys. Rev. Lett., 67, 3328
\bibitem[van Gorkom \etal(1989)]{vang89} van Gorkom, J. H., Knapp, G. R., Ekers, R. D., Ekers, D. D., Laing, R. A., \& Polk, K. S. 1989, \aj, 97, 708 
\bibitem[van Langevelde \etal(1995)]{vanlang95} van Langevelde, H. J., van Dishoeck, E. F., Sevenster, M. N., \& Israel, F. P. 1995, \apj, 448, L123
\bibitem[Vermeulen \etal(2003)]{ver03} Vermeulen, R. C., Pihlstr\"{o}m, Y. M., Tschager, W., \etal\ 2003, \aap, 404, 861
\bibitem[White \& Becker(1992)]{white92} White, R. L. \& Becker, R. H. 1992, \apjs, 79, 331
\bibitem[Wilkinson \etal(1994)]{wilkinson94} Wilkinson, P. N., Polatidis, A. G., Readhead, A. C. S., Xu, W., \& Pearson, T. J. 1994, \apj, 432, L87  
\bibitem[Wills \& Lynds(1978)]{wills78} Wills, D. \& Lynds, R. 1978, \apjs, 36, 317 
\bibitem[Wolfe \& Burbidge(1975)]{wolfe75} Wolfe, A. M. \& Burbidge, G. R. 1975, \apj, 200, 548 
\bibitem[Wright, Ables \& Allen(1983)]{wright83} Wright, A. E., Ables, J. G., \& Allen, D. A. 1983, \mnras, 205, 793
\bibitem[Wright(2006)]{wright06} Wright, E. L. 2006, \pasp, 118, 117 % cosmo calculator
\bibitem[Xiang \etal(2005)]{xiang05} Xiang, L, Dallacasa, D., Cassaro, P., Jiang, D., \& Reynolds, C. 2005, \aap, 434, 123 
\bibitem[Xu \etal(1995)]{xu95} Xu, W., Readhead, A. C. S., Pearson, T. J., Polatidis, A. G., \& Wilkinson, P. N. 1995, \apjs, 99, 297 
\bibitem[Yan \etal(2012)]{yan12} Yan, T., Stocke, J. T., Darling, J., \& Hearty, F. 2012, \aj, 144, 124 
\bibitem[Yan \etal(2016)]{yan16} Yan, T., Stocke, J. T., Darling, J., Momjian, E., Sharma, S., \& Kanekar, N. 2016, \aj, 151, 74
\end{thebibliography}
\end{document}